\RequirePackage{fix-cm}
\documentclass[11pt,a4paper]{article}

\usepackage{amssymb}
\usepackage{ulem}
\usepackage{cancel}
\usepackage{xcolor}
\usepackage[utf8]{inputenc}
\usepackage{amsopn}
\usepackage{bm}
\usepackage{arydshln}
\usepackage{amssymb}
\usepackage{amsthm}
\usepackage[listofformat=parens]{subfig}
\usepackage{float}
\usepackage{hyperref}
\usepackage{stmaryrd}
\usepackage{gensymb}
\usepackage{empheq}
\usepackage[titletoc]{appendix}
\allowdisplaybreaks
\usepackage{chngcntr}
\usepackage{apptools}
\usepackage{caption}
\usepackage{setspace}
\usepackage{stmaryrd}
\usepackage{guit}
\usepackage{mathtools}
\usepackage{graphicx}
\usepackage{psfrag}
\usepackage{pgfplots}
\pgfplotsset{compat=1.17}
\usepackage{float}
\usepackage{bm}
\usepackage{xcolor}
\usepackage{tikz}
\usepackage{lipsum}
\usepackage{amsfonts}
\usepackage{graphicx}
\usepackage{epstopdf}
\usepackage{algorithm}
\usepackage[noend]{algpseudocode}

\newcommand{\llbrace}{\lbrace\hspace{-0.15cm}\lbrace}
\newcommand{\rrbrace}{\rbrace\hspace{-0.15cm}\rbrace}

\begin{document}
	
	\title{Level set-fitted polytopal meshes with application to structural topology optimization}
	
	\author{Nicola Ferro$^1$, Stefano Micheletti$^1$, Nicola Parolini$^1$,\\ Simona Perotto$^1$, Marco Verani$^1$, Paola Francesca Antonietti$^1$}
	\maketitle
	
	\begin{center}
		{\small
			$^1$
			MOX -- 
			Dipartimento di Matematica\\
			Politecnico di Milano\\
			Piazza L. da Vinci, 32, I-20133 Milano, Italy
		}
	\end{center}
	\date{}
	\maketitle
	
	\begin{abstract}
		We propose a method to modify a polygonal mesh in order to fit the zero-isoline of a level set function by extending a standard body-fitted strategy to a tessellation with arbitrarily-shaped elements. 
		The novel level set-fitted approach, in combination with a Discontinuous Galerkin finite element approximation, provides an ideal setting to model physical problems characterized by embedded or evolving complex geometries, since it allows us skipping any mesh post-processing in terms of grid quality.
		The proposed methodology is firstly assessed on the linear elasticity equation, by verifying the approximation capability of the level set-fitted approach when dealing with configurations with heterogeneous material properties. Successively, we combine the 
		level set-fitted methodology with a minimum compliance topology optimization technique, in order to deliver optimized layouts exhibiting crisp boundaries and reliable mechanical performances. An extensive numerical test campaign confirms the effectiveness of the proposed method.
		\\[3mm]
	\end{abstract}
	
	\section{Introduction}
	\label{sec:introduction}
	
	The level set method is recognized as a versatile computational tool for solving a wide range of problems in applied science and engineering, constituting a powerful framework for the efficient and accurate representation of complex boundaries and evolving interfaces~\cite{osher2001}. A level set method embeds the contour of the considered geometry inside a generic domain as the isocontour of a signed-distance function. 
	Such an implicit representation -- often also referred to as an immersed boundary approach -- offers various advantages over traditional boundary tracking methods (such as the explicit conforming meshing~\cite{perotto2022}), allowing one to easily handle topological changes and facilitating the incorporation of diverse physical characteristics into different portions of the domain. These good features justify the  adoption of a level set approach in 
	a huge variety of fields that require an accurate tracking of complex evolving geometries, including computational fluid dynamics, image processing, computer graphics, biomedical engineering, and structural mechanics (see, e.g.,~\cite{gibou2007,samson2000,kim2021,vanDijk2013}).
	
	The numerical discretization of a level set approach in a finite element framework has been analyzed in depth in the literature, e.g., we refer the reader to~\cite{gibou2018}.
	The main issue in such a context remains 
	the accurate tracking of the interface embedded in the level set method since this is strictly connected to the adopted computational mesh. 
	In general, it is advisable to choose a tessellation that balances accuracy and computational cost.
	For instance, an excessively coarse or a poorly aligned grid with respect to the isoline of the level set function may lead to inaccurate results, especially close to the interface~\cite{zhao2014,sheu2011}.
	As possible remedies, 
	body-fitted approaches and mesh adaptation strategies have been extensively adopted in different fields, as, for instance, in~\cite{elaouad2022,antepara2020,cortellessa2023}. Nevertheless, body-fitted approaches may require a careful mesh manipulation and local or global remeshing operations to guarantee a suitable grid quality, especially in the presence of triangular tessellations~\cite{zhuang2021,allaire2011}. 
	Such a fine-tuning of the mesh may lead to a considerable computational overhead and may eventually deteriorate the performance of the overall method.
	
	To limit the post-processing associated with a body-fitted approach for a level set model, we propose a new algorithm which exploits polygonal meshes in terms of flexibility of the element shape. 
	In particular, we introduce a fitting scheme 
	to appropriately cut polygonal elements for the generation of level set-conforming computational grids. \\
	To test in practice the approximation capabilities of the proposed level-set approach, we employ a polygonal discontinuous Galerkin (PolyDG) discretization method. Indeed, it has been proved in~\cite{CangianiDongGeorgoulisHouston_2017,CangianiGeorgoulisHouston_2014} 
	that PolyDG approximations can support arbitrarily-shaped mesh elements with mild regularity requirements (e.g., polygons with an unbounded number of possibly degenerate edges/faces), such as the elements returned by a levet set-fitted approach. \\
	A preliminary successful verification of the novel method onto a standard linear elasticity equation framework 
	paves the way to more challenging applications. Specifically, we address the problem of structural topology optimization, where the accurate resolution of the interface between void and material plays a key role~\cite{sigmund2007,vanDijk2013}. In particular, the novel fitting algorithm enhances the geometrical description of the structure under optimization, while increasing the accuracy of the linear elasticity analysis at the base of the method. 
	
	Body-fitted approaches in a topology optimization setting can be found in the recent literature, mainly confined to triangular or quadrilateral tessellations~\cite{li2022,nardoni2022,zhuang2022,andreasen2020,yamasaki2011}, and resort to localized remeshing operations to improve the grid quality~\cite{kuci2021,zhuang2021}, or employ an eXtended Finite Element approach, which enriches the standard discrete function spaces~\cite{villanueva2014,sharma2017}. Concerning generic polygonal tessellations, only few contributions on topology optimization are available, starting from the seminal papers~\cite{Talischi2009,talischi2012}, and in combination with different discretization schemes, such as in~\cite{wei2020} where the authors focus on a radial basis function-based topology optimization approach, or in~\cite{antonietti2017,tran2023} where a Virtual Element Method is adopted.
	Finally, the reader is referred to other literature contributions where polygonal grids are employed in a topology optimization process, although the trimming procedure relies on a posteriori case-by-case evaluation~\cite{nguyen2022,nguyen2019}, or on grid adaptation techniques~\cite{hoshina2018,chau2018}.
	
	The paper is organized as follows. Section~\ref{sec:fittedmesh} introduces the preliminary concepts on the level set method and details the fitting methodology. Such approach is exemplified on the linear elasticity case and numerically tested on a benchmark configuration. In Section~\ref{sec:topopt}, we briefly present the minimum compliance problem in level set-topology optimization and we sketch the algorithm to couple topology optimization and the mesh fitting procedure. The proposed workflow is verified on three test cases. Finally, some conclusions and some possible future developments are gathered in Section~\ref{sec:conclusions}.
	
	\section{Level set-fitted polygonal discretization}
	\label{sec:fittedmesh}
	
	The standard level set method employs a signed-distance function $\varphi: \Omega \rightarrow \mathbb{R}$ that partitions the bounded Lipschitz domain $\Omega \subset \mathbb R^2$ in different portions. The one-dimensional curve $\mathcal{C} = \{ {\bm x} \in \Omega : \varphi({\bm x}) = d \}$ represents the interface between the two parts of the domain, identified by $\{ {\bm x} \in \Omega : \varphi({\bm x}) > d \}$ and $ \{ {\bm x} \in \Omega : \varphi({\bm x}) < d \}$, with $d \in \mathbb R$ a selected threshold. This modeling expedient allows us to immerse a generic geometry, whose boundary is represented by the $d$-isoline of $\varphi$, inside the bounding domain $\Omega$. Thus, the embedded geometry can enter implicitly into the physical problem under consideration, can be subject to an evolution, and can undergo topological changes, through the continuous variable $\varphi$.
	
	An accurate numerical discretization of the level set function is expected to take into account the geometry of the $d$-isoline to properly reproduce the associated shape and curvature. In particular, a mesh-based discretization should be aligned and/or refined in correspondence with the curve $\mathcal C$ in order to properly resolve the interface and to reduce the approximation errors.
	
	This section aims at verifying that the perks led by the level set-fitted approach in different practical contexts (see, e.g.,~\cite{elaouad2022,antepara2020,cortellessa2023}) are preserved when dealing with a generic polytopal tessellation of the domain. Though the proposed fitting procedure is general and can be employed in combination with various discretization schemes, this investigation is carried out in the setting of polygonal discontinuous finite elements, in order to benefit also of the flexibility and the robustness of the discontinuous Galerkin (DG) method when dealing with heterogeneous data.
	
	\subsection{The level set-fitted methodology on polygonal meshes}
	\label{sec:fittedmesh_method}
	We leverage the flexibility of polygonal tessellations in a polygonal discontinuous Galerkin  framework (PolyDG~\cite{AntoniettiGianiHouston_2013,CangianiGeorgoulisHouston_2014,CangianiDongGeorgoulis_2017,collins2016,facciola2021}) in order to generate level set-fitted computational grids, which do not require global/local mesh quality-oriented adjustments. This capability is inherent to the DG method for polygonal tessellations, which delivers reliable and accurate solutions even in the presence of non-standard-shaped (e.g., non-convex) polygonal elements, with possibly degenerate edges or elements characterized by an unbounded number of faces/edges~\cite{CangianiDongGeorgoulisHouston_2017}.
	
	In this framework, we introduce a polytopic mesh $\mathcal{T}_h$ composed by general polygons featuring minimal requirements, as indicated in~\cite{CangianiDongGeorgoulisHouston_2017}.
	In the set $\mathcal{F}_h^{i}$, we collect  the (non-empty) intersections between neighbouring elements in $\mathcal{T}_h$, while in $\mathcal{F}^{b}_h$ we gather the (non-empty) intersections between mesh elements and the domain boundary $\partial \Omega$. We set $\mathcal{F}_h=\mathcal{F}_h^{i}\cup\mathcal{F}^{b}_h$.
	\\
	To approximate the level set function $\varphi$, we consider the finite-dimensional space ${V}_h= \{ v_h \in L^2(\Omega) : v_h |_\kappa \in \mathbb{P}_{p}(\kappa), \ \forall \kappa \in \mathcal{T}_h \}$, where $\mathbb{P}_{p}(\kappa)$ is the space of the polynomials of total degree less than or equal to $p \ge 1$, for any $\kappa\in\mathcal{T}_h$. With this formalism, we have that
	$$
	\varphi_h |_\kappa({\bm x}) = \sum_{i = 1}^{N_{\rm dof}^{\rm loc}} c_i \Phi_i({\bm x}), \quad \forall \bm x \in \kappa, \quad \forall \kappa \in \mathcal{T}_h,
	$$
	is the PolyDG approximation of the continuous level set function on the generic element $\kappa$, where $\{c_i\}_{i = 1}^{N_{\rm dof}^{\rm loc}}$ denotes the set of the expansion coefficients of the discrete function with respect to the local basis of the space $ \mathbb{P}_{p}(\kappa)$, with $N_{\rm dof}^{\rm loc} = (p + 1)(p+2)/2$ the associated dimension.
	
	Without loss of generality, in the following we select the level-set threshold $d$ equal to $0$. In order to make the polygonal mesh fitted with respect to the zero-isoline of $\varphi_h$, we compute the set $\mathcal{C}_h = \{ {\bm x} \in \Omega : \varphi_h({\bm x}) = 0 \}$, which locates the zero-curve of the discrete function and is instrumental to drive the mesh edge insertion procedure. We can define set $\mathcal{C}_h$ as the union of the elementwise zero-isolines $\ell_h^\kappa$, i.e.,
	
	$$
	\mathcal{C}_h = \bigcup_{\kappa \in \mathcal{T}_h} \{ \ell_h^\kappa \subset \kappa : \varphi_h|_\kappa({\bm x}) = 0, \quad \forall  {\bm x} \in \ell_h^\kappa \}.
	$$
	Throughout the paper, we consider linear finite elements (i.e., $p = 1$). As a consequence, local zero-isolines $\ell_h^\kappa$ reduce either to the empty set or to a line segment. The extension to $p > 1$ can be easily implemented by projecting $\varphi_h$ onto the finite element space of local degree $1$. 
	
\begin{figure}[hbt]
		\centering
		\includegraphics[width=0.95\textwidth, keepaspectratio]{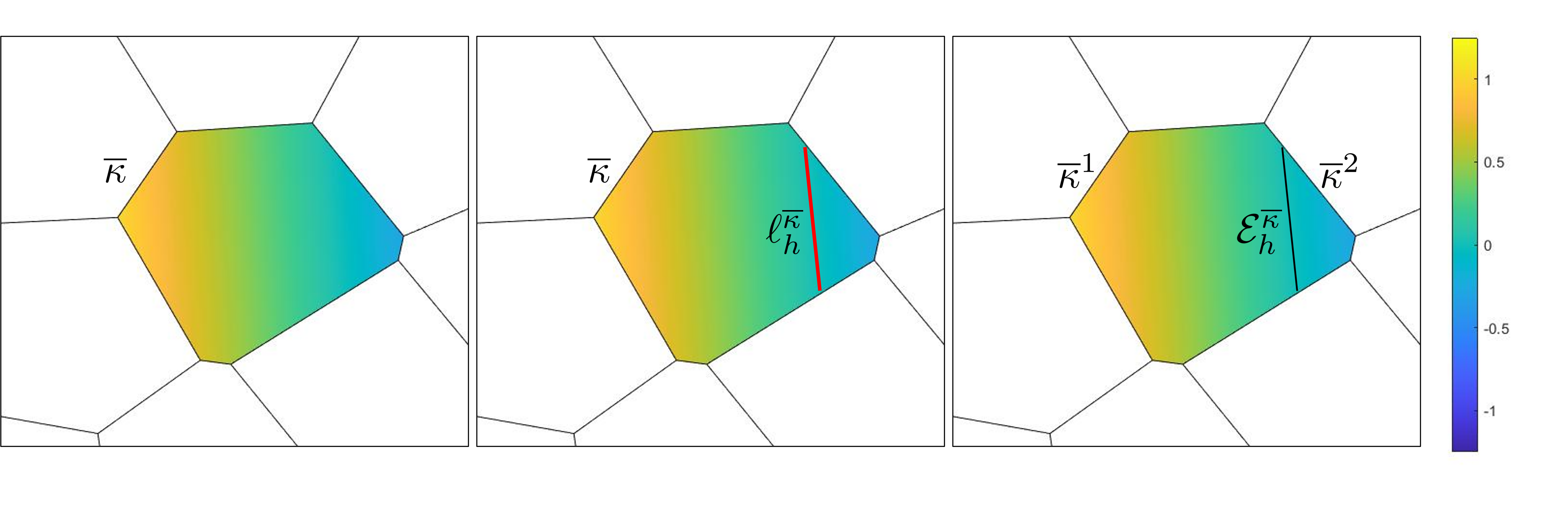}
		\caption{Elementwise representation of the level set-fitted  Algorithm~\ref{cutalgorithm}.}
		\label{fig:cut_procedure}
	\end{figure}

	The level set-fitted mesh generation procedure inserts a new mesh edge, $\mathcal{E}_h^{\overline{\kappa}}$, in each element $\overline{\kappa}$ that satisfies $\ell_h^{\overline{\kappa}} \neq \emptyset$, namely for each element that is crossed by a zero-isoline (see~Figure~\ref{fig:cut_procedure}). Thus, the considered element $\overline{\kappa}$ is split into two new neighbouring elements, $\overline{\kappa}^1$ and $\overline{\kappa}^2$, which share the newly generated face $\mathcal{E}_h^{\overline{\kappa}}$, coinciding with $\ell_h^{\overline{\kappa}}$. The procedure is described in Algorithm~\ref{cutalgorithm}, where we provide the essential steps to compute the fitted mesh. In particular, the {\tt fitMesh} routine loops through all the polygonal elements and, via function {\tt edgeInsert}, inserts a new mesh edge if the local zero-isoline exists. When the loop is terminated, the cutline is assembled and the mesh is fitted by updating elements, edges and connectivity in the mesh structure (subroutines {\tt cutLine} and {\tt constructMesh}, respectively).
	
	\begin{algorithm}[H]
		\caption{{\tt fitMesh}}\label{cutalgorithm}
		{\bf Input} :  $\mathcal{T}_h$, $\varphi_h$
		\begin{algorithmic}[1]
			\State Set: $\mathcal{C}_h = $[ ], ${\tt i} = 0$, $\mathcal{E}_h^\kappa=\mathbf{0}$,
			\vspace{2mm}
			\While {${\tt i} < \#\mathcal{T}_h$}
			\vspace{2mm}
			\State $\kappa = ${\tt extractElement}($\mathcal{T}_h$, ${\tt i}$);
			\vspace{2mm}
			\State $\ell_h^\kappa = ${\tt isoline}($\kappa$, $\varphi_h|_\kappa$);
			\vspace{2mm}
			\If{$\ell_h^\kappa \neq \emptyset$ }
			\vspace{2mm}
			\State $\mathcal{E}_h^\kappa[{\tt i}]$ = {\tt edgeInsert}($\ell_h^\kappa$);
			\EndIf
			\State {\bf end if}
			\vspace{2mm}
			\State ${\tt i} = {\tt i}+{\tt 1}$;
			\vspace{2mm}
			\EndWhile
			\State {\bf end while}
			\vspace{2mm}
			\State $\mathcal{C}_h = ${\tt cutLine}($\mathcal{E}_h^\kappa$);
			\vspace{2mm}
			\State $\mathcal{T}_h^{\rm fit} = ${\tt constructMesh}($\mathcal{T}_h$, $\mathcal{C}_h$);
		\end{algorithmic}
		{\bf Output}: $\mathcal{T}_h^{\rm fit}$
	\end{algorithm}
	
	We remark that the methodology in Algorithm~\ref{cutalgorithm} can be extended to three-dimensional polytopal meshes. In particular, functions {\tt isoline}, {\tt edge\-Insert} and {\tt cutLine} should be modified into {\tt isocontour}, {\tt planeInsert} and {\tt cutSurf}, respectively, to implement elemental cuts through planes. Nevertheless, such extension deserves specific adjustments that go beyond the scope of this work. 
	
	\subsection{Application to the linear elasticity system}
	\label{sec:linearelasticity}
	As a paradigmatic example, we apply the level set-fitted procedure to the mechanical analysis of a two-dimensional linear elastic body $\Omega$. The boundary of the domain, $\partial \Omega$, is decomposed into the portions $\Gamma_D$, $\Gamma_N$, and $\Gamma_F$, where Dirichlet, non-homogeneous and homogeneous Neumann conditions are applied, respectively, so that $\partial \Omega = \Gamma_D \cup \Gamma_N \cup \Gamma_F$. The linear elasticity equation for the displacement, $\bm u$, reads~\cite{gould}
	\begin{equation} \label{eq:linear_elasticity}
		\begin{cases}
			- \nabla \cdot \bm \sigma = \bm{f} & \textrm{ in }\Omega, \\
			\bm u = \mathbf{0} & \textrm{ on } \Gamma_D, \\
			\bm{\sigma}  \, \bm n = \bm g & \textrm{ on }\Gamma_N , \\
			\bm{\sigma}  \, \bm n = \bm 0 & \textrm{ on }\Gamma_F,
		\end{cases}
	\end{equation}
	where the function $\bm{f}$ represents the body force applied to the system and $\bm g$ is the traction exerted on the boundary portion $\Gamma_N$.
	Concerning the constitutive law that relates the stress field $\bm \sigma$ to the strain $\bm \epsilon (\bm u) = (\nabla \bm u + \nabla \bm u^T)/2$, we consider an isotropic linear law, i.e.,
	$$
	\bm \sigma  = \mathbb{D} \bm{\epsilon}(\bm u) = \lambda \nabla \cdot \bm u \bm\, I  + 2 \mu \,\bm \epsilon(\bm u) \quad \textrm{ in }\Omega,
	$$
	with $\mathbb{D}$ the stiffness tensor. Such quantity -- possibly varying in $\Omega$ in correspondence with different materials -- depends on the Lam\'e coefficients, $\lambda$ and $\mu$, given by the expressions
	\begin{equation}\label{Lcoef}
		\lambda=\dfrac{E\nu}{(1+\nu)(1-2\nu)},\quad \mu=\dfrac{E}{2(1+\nu)},
	\end{equation}
	with $E$ and $\nu$ the Young modulus and the Poisson ratio of the material, respectively.
	
	The weak formulation of problem~\eqref{eq:linear_elasticity} is: {\it find $\bm u\in  [H^{1}_{\Gamma_D}(\Omega)]^2$ such that}
	\begin{equation}\label{eq:linear_weak}
		\mathcal{A}^e(\bm u, \bm v) = (\bm f, \bm v)_\Omega + (\bm g, \bm v)_{\Gamma_N} \quad  \forall\, \bm v \in  [H^{1}_{\Gamma_D}(\Omega)]^2,
	\end{equation}
	with $H^{1}_{\Gamma_D}(\Omega)$ the standard scalar Sobolev space of index $1$ of real-valued functions defined on $\Omega$ with null trace on $\Gamma_D$, and where, for any $\bm w,\bm z \in [H^{1}_{\Gamma_D}(\Omega)]^2$, we set \begin{equation}\label{eq:el_term}
		\mathcal{A}^e(\bm w, \bm z) = 
		( \bm{\sigma}(\bm w) , \bm{\epsilon}(\bm z))_\Omega = ( 2 \mu \bm \epsilon (\bm w) , \bm{\epsilon}(\bm z))_\Omega + (\lambda \nabla \cdot \bm w, \nabla \cdot \bm z)_\Omega,
	\end{equation}
	with $(\cdot, \cdot)_\omega$ the standard $L^2(\omega)$-inner product in $\omega \subseteq \Omega$.
	
	The PolyDG discretization of problem \eqref{eq:linear_weak} requires the definition of some operators. Following~\cite{Arnoldbrezzicockburnmarini2002}, upon considering sufficiently piecewise smooth scalar-, vector- and tensor-valued fields $\psi$, $\bm{v}$ and $\bm{\tau}$, respectively,
	we define the jumps ($\llbracket \cdot \rrbracket$) and the averages ($\llbrace \cdot \rrbrace$) on each interior face $F\in\mathcal{F}_h^{i}$ shared by the neighbouring elements $\kappa^{+}$ and $\kappa^{-} \in \mathcal{T}_h$, as
	$$
	\begin{array}{ll}
		\llbracket\psi\rrbracket =  \psi^+\bm{n}^++\psi^-\bm{n}^-,
		&
		\llbrace\psi \rrbrace  = \dfrac{\psi^++\psi^-}{2}
		\\[3mm]
		\llbracket\bm{v}\rrbracket  = \bm{v}^+\otimes\bm{n}^++\bm{v}^-\otimes\bm{n}^-,
		& 
		\llbrace\bm{v}\rrbrace   = \dfrac{\bm{v}^++\bm{v}^-}{2},
		\\[3mm]
		\llbracket\bm \tau \rrbracket_{\bm n} = \bm{\tau}^+\cdot\bm{n}^++\bm{\tau}^-\cdot\bm{n}^-,
		&
		\llbrace\bm{\tau}\rrbrace  = \dfrac{\bm{\tau}^++\bm{\tau}^-}{2},
	\end{array}
	$$
	with $\bm v^{\pm} \otimes \bm n^\pm = \bm v^\pm (\bm n^\pm)^T$, superscripts $+$ and $-$ denoting the function trace on $F$, taken within $\kappa^+$ and $\kappa^-$, respectively and with $\bm{n}^+$ and $\bm{n}^-$ the unit outer normal vectors to $F$ with respect to $\kappa^+$ and $\kappa^-$.  
	Along the boundary faces, $F\in\mathcal{F}_h^{b}$, the jumps and the averages are defined by
	$$
	\llbracket\psi\rrbracket = \psi\bm{n},\
	\llbrace\psi \rrbrace = \psi,\
	\llbracket\bm{v}\rrbracket= \bm{v}\otimes\bm{n},\
	\llbrace\bm{v}\rrbrace= \bm{v},\
	\llbracket\bm{\tau}\rrbracket_{\bm n} =\bm{\tau}\cdot\bm{n},\
	\llbrace\bm{\tau}\rrbrace= \bm{\tau}.
	$$  
	After splitting the boundary faces into the Dirichlet, $\mathcal{F}^{D}_h$, and Neumann, $\mathcal{F}^{N}_h$, subsets that are conformal with respect to $\Gamma_D$ and $\Gamma_F \cup \Gamma_N$, respectively the Symmetric Interior Penalty PolyDG approximation of problem \eqref{eq:linear_weak} can be formulated~\cite{Riviere}: {\it find $\bm u_h\in  \bm{U}_h = [V_h]^2$ such that}
	\begin{equation}\label{eq:linear_dg}
		\mathcal{A}^e_h(\bm u_h, \bm v_h) = (\bm f, \bm v_h)_{\mathcal{T}_h} + (\bm g, \bm v_h)_{\mathcal{F}_h^N} \quad  \forall\, \bm v_h \in \bm U_h,
	\end{equation}
	where the bilinear form $\mathcal{A}_h^e: \bm{U}_h \times \bm{U}_h \rightarrow \mathbb{R}$ is given by 	 
	\begin{equation}\label{eq:linear_discrete}
		\begin{array}{lcl}
			& \phantom{.} & \mathcal{A}_h^e(\bm w_h,\bm z_h) 
			=
			(2 \mu \bm \epsilon_h(\bm w_h) , \bm{\epsilon}_h(\bm z_h))_{\mathcal{T}_h} 
			+
			(\lambda \nabla_h \cdot \bm w_h, \nabla_h \cdot \bm z_h)_{\mathcal{T}_h}
			\\[3mm]
			&-&( \llbrace 2 \mu \bm \epsilon_h(\bm w_h)\rrbrace,  \llbracket \bm z_h  \rrbracket)_{\mathcal{F}_h^i\cup\mathcal{F}_h^D}
			-( \llbracket \bm w_h  \rrbracket, \llbrace 2 \mu \bm \epsilon_h(\bm z_h)\rrbrace)_{\mathcal{F}_h^i\cup\mathcal{F}_h^D}
			\\[3mm]
			&-&( \llbrace \lambda \nabla_h \cdot \bm w_h\rrbrace,  \llbracket \bm z_h  \rrbracket_{\bm n})_{\mathcal{F}_h^i\cup\mathcal{F}_h^D}
			-( \llbracket \bm w_h  \rrbracket_{\bm n}, \llbrace \lambda \nabla_h \cdot \bm z_h\rrbrace)_{\mathcal{F}_h^i\cup\mathcal{F}_h^D}
			\\[3mm]
			&+&  (\eta_\mu\,\llbracket \bm w_h  \rrbracket, \llbracket \bm z_h  \rrbracket)_{\mathcal{F}_h^i\cup\mathcal{F}_h^D}
			+ (\eta_\lambda\,\llbracket \bm w_h  \rrbracket_{\bm n}, \llbracket \bm z_h  \rrbracket_{\bm n})_{\mathcal{F}_h^i\cup\mathcal{F}_h^D}
			\quad \forall \bm w_h,\bm z_h \in \bm U_h,
		\end{array}
	\end{equation}
	where $(\cdot,\cdot)_{\mathcal{T}_h} = \sum_{\kappa\in \mathcal{T}_h} (\cdot,\cdot)_\kappa$ and $(\cdot,\cdot)_{\mathcal{F}_h^I \cup \mathcal{F}_h^D} = \sum_{F\in \mathcal{F}_h^I \cup \mathcal{F}_h^D} (\cdot,\cdot)_F$ 
	is the adopted compact notation for the inner products. In the PolyDG formulation, we replace the differential operator $\nabla$ with the broken counterpart $\nabla_h$, so that $\nabla_h \theta_h := \nabla \theta_h|_\kappa$ and $\nabla_h \cdot  \bm \theta_h := \nabla \cdot \bm \theta_h|_\kappa$ for any $\kappa \in \mathcal{T}_h$, while $\bm \epsilon_h(\bm \theta_h) := (\nabla_h \bm \theta_h + \nabla_h \bm \theta_h^T)/2$, for generic functions $\theta_h \in V_h$ and $\bm \theta_h \in \bm U_h$.
	\\
	The jump penalization in \eqref{eq:linear_discrete} is imposed through the functions $\eta_\mu:\mathcal{F}_h\rightarrow\mathbb{R}^+$ and $\eta_\lambda:\mathcal{F}_h\rightarrow\mathbb{R}^+$, defined facewise as 
	\begin{equation}\label{eq:penalization_mu}
		\eta_\mu=\sigma_{0, \mu}  
		\begin{cases}
			\underset{\kappa\in\{\kappa_1,\kappa_2\} } \max \mu_\kappa C(p, \kappa, F), & F \in \mathcal{F}_h^i, \, F \subseteq \partial \kappa_1 \cap \partial \kappa_2, \\
			\mu_\kappa C(p, \kappa, F), &  F\in\mathcal{F}_h^D,  \, F \subseteq \partial \kappa \cap \Gamma_D,
		\end{cases}
	\end{equation}
	\begin{equation}\label{eq:penalization_lambda}
		\eta_\lambda=\sigma_{0, \lambda}  
		\begin{cases}
			\underset{\kappa\in\{\kappa_1,\kappa_2\} } \max \lambda_\kappa C(p, \kappa, F), & F \in \mathcal{F}_h^i, \, F \subseteq \partial \kappa_1 \cap \partial \kappa_2, \\
			\lambda_\kappa C(p, \kappa, F), &  F\in\mathcal{F}_h^D,  \, F \subseteq \partial \kappa \cap \Gamma_D,
		\end{cases}
	\end{equation}
	where $\mu_\kappa = \max_{\bm x \in F \cap \partial \kappa} \mu(\bm x)$ and $\lambda_\kappa = \max_{\bm x \in F \cap \partial \kappa} \lambda(\bm x)$ are the interface material Lam\'e coefficients in the element $\kappa \in \mathcal{T}_h$, $\sigma_{0, \mu}$ and $\sigma_{0, \lambda}$ are (large enough) positive constants to be chosen, and $C(p, \kappa, F)$ is a scaling geometric factor, which turns out to be suited for generic-shaped polygonal elements, according to~\cite{CangianiGeorgoulisHouston_2014}.
	We remark that $\mu_\kappa$ and $\lambda_\kappa$ are simplified into $\mu|_\kappa$ and $\lambda|_\kappa$ in case of piecewise constant material parameters.
	
	\subsubsection{An example with a two-material domain} \label{sec:lshape_validation}
\begin{figure}[hbt]
		\centering
		\includegraphics[width=0.3\textwidth, keepaspectratio]{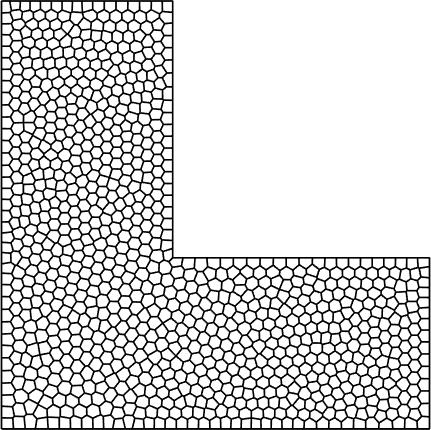} \quad
		\includegraphics[width=0.3\textwidth, keepaspectratio]{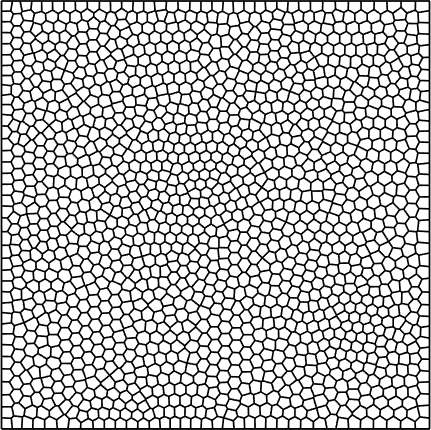} \quad
		\includegraphics[width=0.3\textwidth, keepaspectratio]{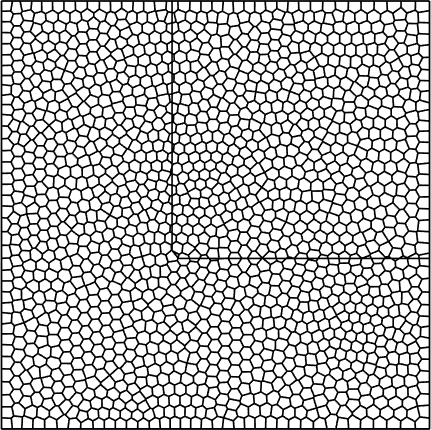}
		\caption{Test case of Section~\ref{sec:lshape_validation} - Example of the considered meshes for test cases i) (left), ii) (center), and iii) (right).
		}
		\label{fig:Lshapes}
	\end{figure}

	We validate the level set-fitted methodology on a benchmark test case in linear elasticity, by considering the structural response of an L-shaped loaded structure. With this regard, we compare three different approaches to model such a layout. In particular, we examine:
	\begin{enumerate}
		\item[i)]
		an L-shaped geometry $\Omega_{1} = (0, 10)^2 \setminus [4, 10]^2$, which is explicitly tessellated with a polygonal mesh $\mathcal{T}_{h, 1}$ (see an example in Figure~\ref{fig:Lshapes}, left);
		\item[ii)]
		an L-shaped configuration, which is embedded through a level set function, $\varphi_L$, in a bounding domain $\Omega_{2} = (0, 10)^2$, discretized through a polygonal mesh $\mathcal{T}_{h, 2}$ (see an example in Figure~\ref{fig:Lshapes}, center);
		\item[iii)]
		an L-shaped configuration, which is embedded through a level set function, $\varphi_L$, in a bounding domain $\Omega_{2}$, meshed with a level set-fitted polygonal grid $\mathcal{T}_{h, 3}$ (see an example in Figure~\ref{fig:Lshapes}, right).
	\end{enumerate}
	In particular, function $\varphi_L:\Omega_2 \rightarrow \mathbb{R}$ is chosen so that
	$$
	\left\{
	\begin{array}{rl}
		\varphi_L(\bm x) < 0 &  \text{ for } \{x > 4, y > 4\},\\
		\varphi_L(\bm x) = 0 &  \text{ for } \{x = 4, y \ge 4\} \cup \{x \ge 4, y = 4\},\\
		\varphi_L(\bm x) > 0 &  \text{ for } \{x < 4\} \cup \{y < 4\},
	\end{array}
	\right.
	$$
	in order to correctly immerse the L-shape inside the domain $\Omega_2$.
	The implicit layouts in ii) and iii) are made comparable with the explicit model in i), by filling the portion $W = \{ \bm x \in \Omega_{2} : \varphi_L(\bm x) < 0 \}$ with a soft material. More precisely, by following a standard ersatz weak material approach~\cite{berggren2012,allaire}, the stiffness tensor is defined as
	\begin{equation}\label{eq:ersatz}
		\mathbb{D} = \mathbb{D}(\bm x) = \left\{
		\begin{array}{rl}
			\mathbb{D}^0 & \bm x \in \Omega_{2} \setminus W, \\
			\gamma \mathbb{D}^0 & \bm x \in  W, 
		\end{array}
		\right.
	\end{equation}
	with $\mathbb{D}^0$ the tensor associated with the chosen material in i), characterized by the Young modulus $E^0$ and the Poisson ratio $\nu^0$, and $\gamma \in (0, 1)$ a softness parameter that tunes the severity of the material discontinuity.
	We remark that the limit $\gamma \rightarrow 0^+$ pushes configurations ii) and iii) towards the material-void explicit configuration in i), by assigning negligible stiffness to the material located in $W$. 
	
	In order to assess the effectiveness of the level set-fitted mesh, we hereby study how the choice of the parameter $\gamma$ and the use of the settings in ii) and iii) 
	impact on the accuracy of the finite element solution with respect to the standard approach i).
	This analysis is carried out by setting in \eqref{eq:linear_elasticity} $\bm f = \bm 0$, $\Gamma_D = \{\bm x \in \partial\Omega_j : y = 10\}$, $\Gamma_N = \{\bm x \in \partial \Omega_j : x = 10, y \in [0, 4]\}$, $\Gamma_F = \partial \Omega_j \setminus (\Gamma_D \cup \Gamma_N)$, for $j = 1, 2$, and the upward traction $\bm g = [0, -6(y - 1.5)(y - 2.5) \mathbb{I}_{[1.5, 2.5]}(y)]^T$, with $\mathbb{I}_{[a, b]}$ the indicator function associated with the interval $[a, b]$. Finally, the chosen material is characterized by $E^0 = 1$ and $\nu^0 = 0.3$.
	\\
	We consider two sets of meshes: the coarse meshes $\mathcal{T}_{h, j}^c$, which are characterized by $1920$, $3000$ and $3078$ polygonal elements, for $j = 1, 2, 3$, respectively; the fine meshes, $\mathcal{T}_{h, j}^f$, consisting of $6400$, $10000$ and $10142$ polygons, for $j = 1, 2, 3$, respectively. In particular, the mesh cardinality in i) is always chosen so that the discretized L-shaped domain is tessellated with elements of the same characteristic size as those in ii), while meshes $\mathcal{T}_{h, 3}^c$ and $\mathcal{T}_{h, 3}^f$ are obtained from the non-fitted counterpart in ii) with a $2.6\%$ and $1.4\%$ cardinality increase due to the fitting operations in Algorithm~\ref{cutalgorithm}. 
	\begin{figure}[hbt]
		\centering
		\includegraphics[width=0.475\textwidth, keepaspectratio]{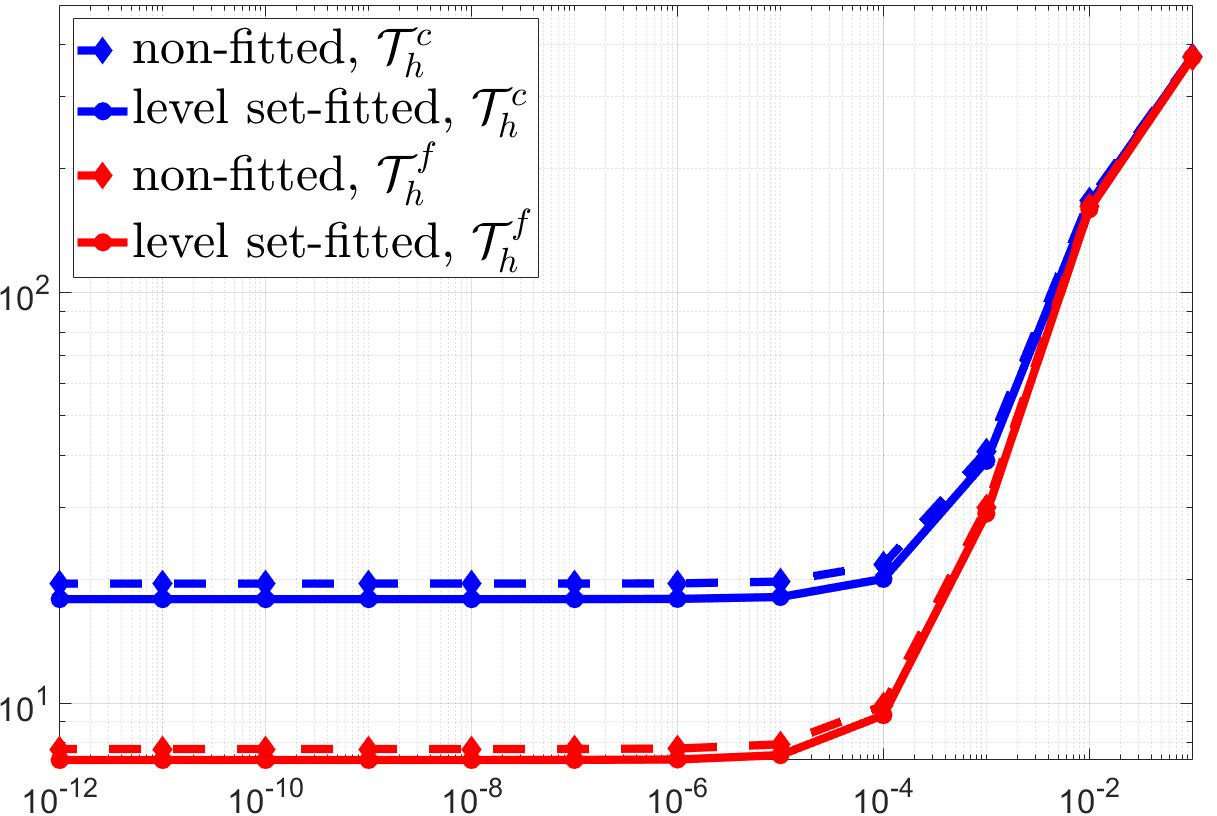}
		\includegraphics[width=0.475\textwidth, keepaspectratio]{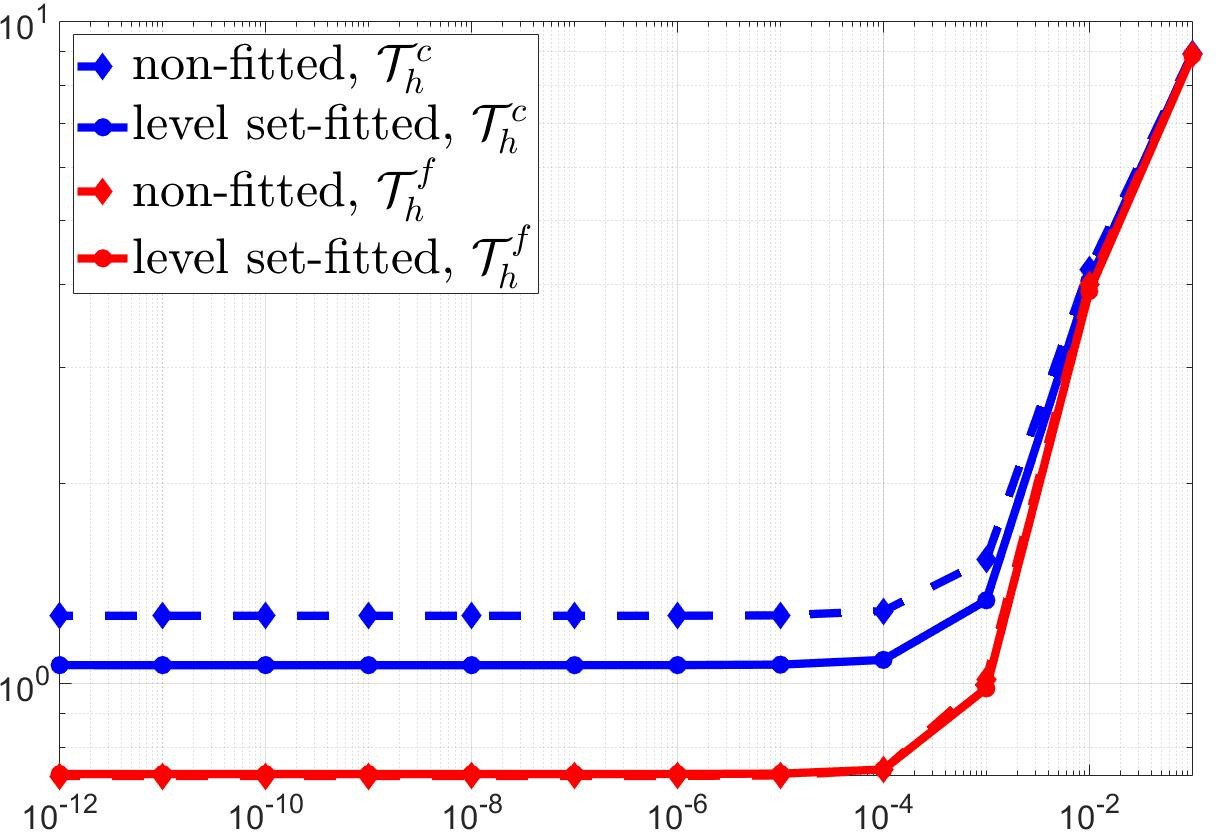}
		\caption{Test case of Section~\ref{sec:lshape_validation} - Computed $L^2(\Omega_1)$-error associated with the displacement field (left) and the von Mises stress (right) for meshes $\mathcal{T}_{h, j}^c$ and $\mathcal{T}_{h, j}^f$, with $j = 2$ (non-fitted) and $j= 3$ (level set-fitted), as a function of $\gamma$.}
		\label{fig:Lshapes_convergence}
	\end{figure}
	
\begin{figure}[hbt]
		\centering
		\includegraphics[width=0.4\textwidth, keepaspectratio]{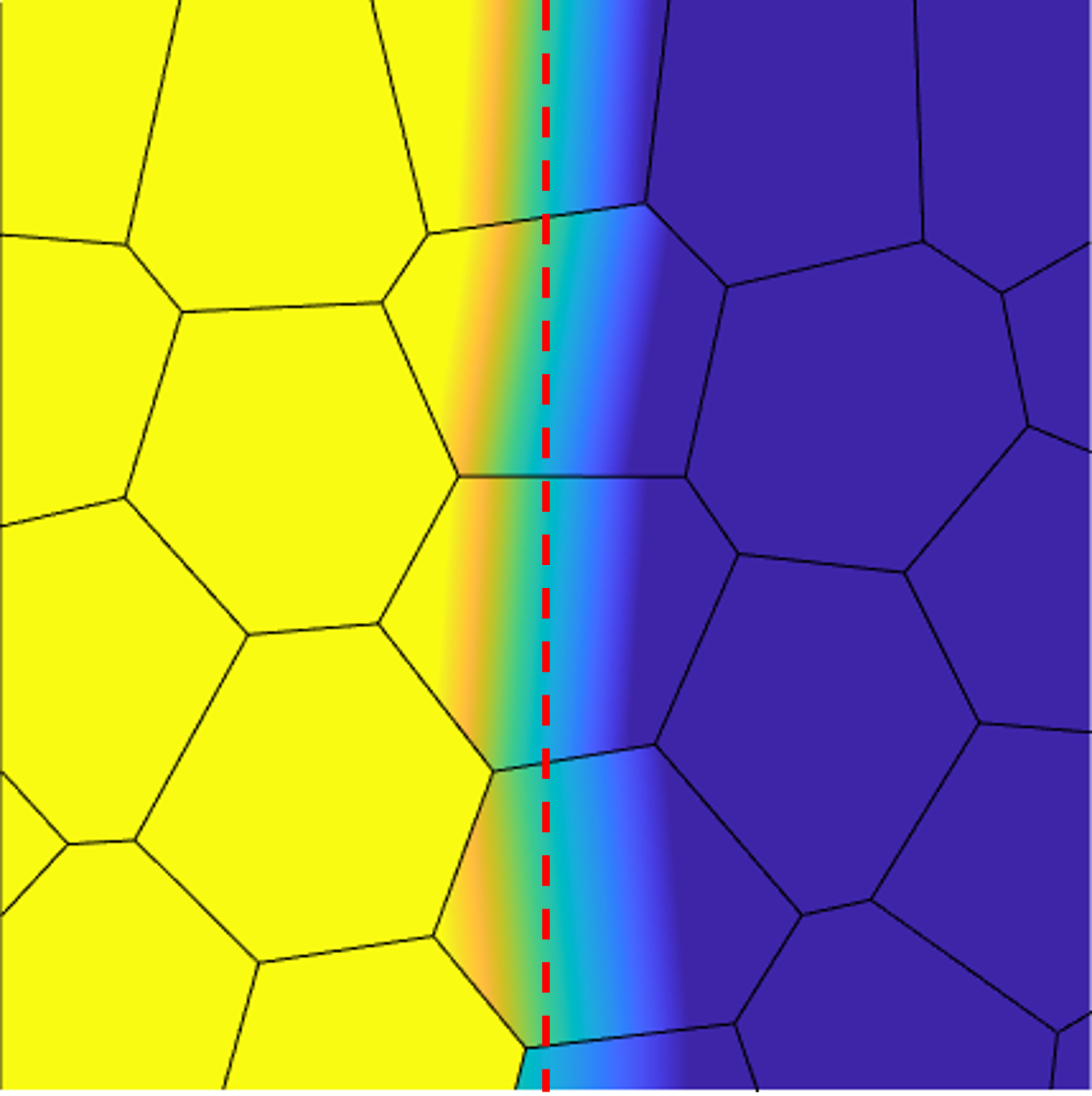} \quad
		\includegraphics[width=0.4\textwidth, keepaspectratio]{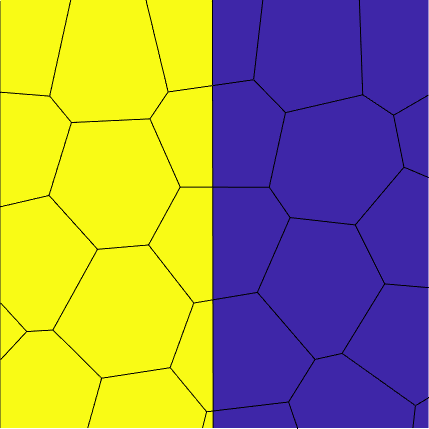}
		\caption{Test case of Section~\ref{sec:lshape_validation} - Enlarged view of the two material-transition in correspondence with the vertical interface for the non-fitted (left) and the level set-fitted (right) mesh.}
		\label{fig:Lshapes_zoom}
	\end{figure}

	In Figure~\ref{fig:Lshapes_convergence}, we show the $L^2(\Omega_1)$-norm of the discretization error associated with the displacement, $\bm u_h$, (left) and the corresponding von Mises stress, $ \sigma_{VM}(\bm u_h)$, (right) as a function of the softness parameter $\gamma$ in \eqref{eq:ersatz}. As a reference solution, we take the continuous finite element approximation, $\bm u_{\rm ref}$, computed on an isotropic triangular discretization, $\mathcal{T}_{h, \rm ref}$, of $\Omega_1$, consisting of $93572$ elements (we refer to~\cite{berggren2012} for a similar analysis). \\
	We can observe that configurations ii) and iii) present the behaviour highlighted in~\cite{berggren2012}, where the errors stagnate around a positive value for $\gamma \lesssim 1$e$-6$. The curves in Figure~\ref{fig:Lshapes_convergence} reach a plateau for small values of $\gamma$, when the error contribution ascribed to the finite element approximation becomes predominant with respect to the one associated with the soft material approximation. It is to notice that the level set-fitted mesh allows us reaching a lower error for both the displacement and the von Mises stress, since avoiding any material approximation along the interface.
	Indeed, the ersazt model may present an additional error if the computational mesh is not aligned with the zero-isoline. Thus, in scenario ii), the weak material expedient may introduce a non-sharp transition of the physical parameters inside an element, as highlighted in Figure~\ref{fig:Lshapes_zoom}, left. On the contrary, the use of a fitted mesh removes such additional error contribution, restoring a conforming and more physical binary representation of the two materials in the domain (see Figure~\ref{fig:Lshapes_zoom}, right). \\
	A comparison between the blue and red lines in Figure~\ref{fig:Lshapes_convergence} shows that the mismatch between scenarios ii) and iii) is noticeable for the coarse mesh, especially for the von Mises analysis (right panel). This behavior is expected since fine meshes dampen the error due to the ersatz approximation by globally increasing the mesh resolution.
	
	The benefits of a level set-fitted mesh are confirmed by comparing the vertical displacement field component along the cut line $r: \{y = x,  3.93 \le x \le  3.95\}$, close to the re-entrant singular corner, 
	for the three selected scenarios and for two different values of $\gamma$. For accuracy reasons, we compute the displacement on meshes $\mathcal{T}_{h, j}^f$, for $j=1$, $2$, $3$.\\
	It turns out that, with respect to the explicit configuration i), the level set-fitted layout in iii) catches the structure response better than scenario ii), due to the possibly inexact localization of the material interface.
	\begin{figure}[h!]
		\centering
		\includegraphics[height=0.45\textwidth, keepaspectratio]{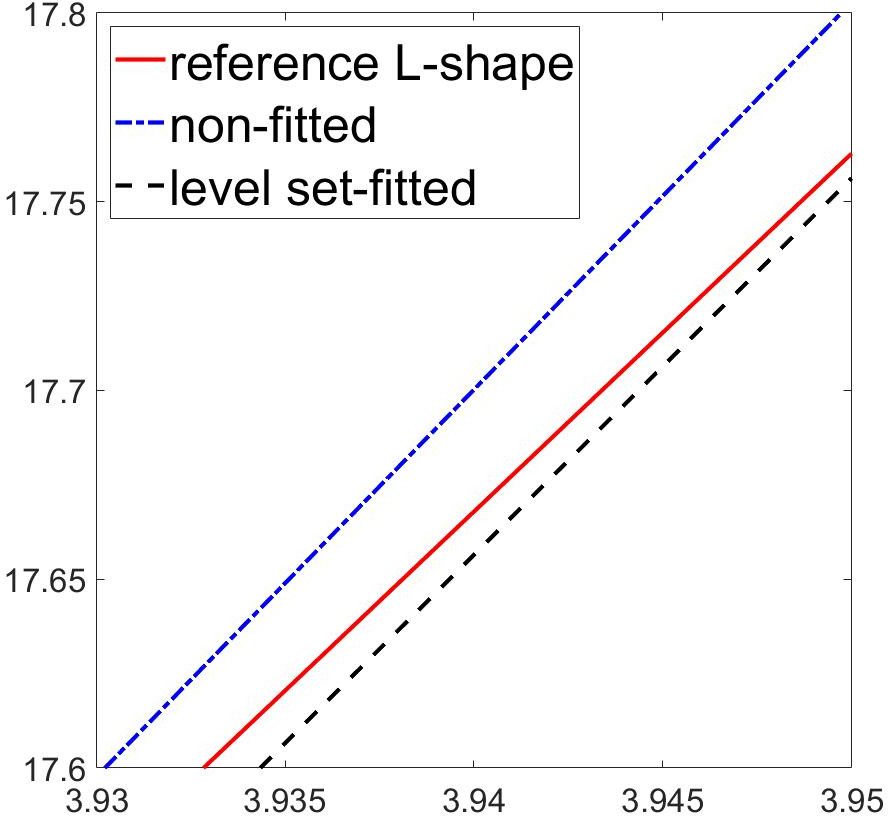} 
		\includegraphics[height=0.45\textwidth, keepaspectratio]{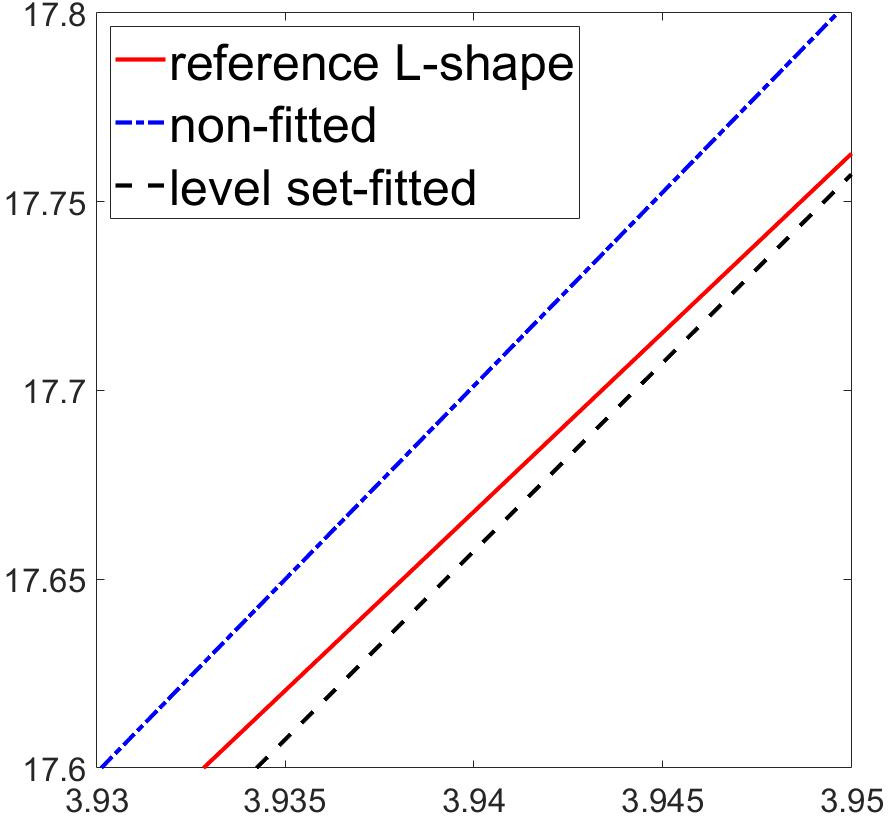}
		\caption{Test case of Section~\ref{sec:lshape_validation} - PolyDG approximation of the vertical displacement field component along the cut line $r: \{y = x,  3.93 \le x \le  3.95\}$ for scenarios i), ii), and iii), for $\gamma = 1$e-$6$ (left) and $\gamma = 1$e-$12$ (right).}
		\label{fig:Lshapes_cutline}
	\end{figure}

	To conclude this assessment, we examine a physical quantity of interest that is relevant for structural applications, namely the static compliance of the structure (i.e., the inverse of the structural stiffness), which coincides with the work done by the external forces, being 
	\begin{equation}\label{eq:compliance}
		l(\bm u_h) = \int_{\Gamma_N} \bm g \cdot \bm u_h ds.
	\end{equation}
	For this purpose, in 
	Table~\ref{tab:compliance} we gather the percentage error on the compliance with respect to the explicit configuration i), namely
	\begin{equation}\label{eq:delta}
		\Delta_\% l (\mathcal{T}_{h,i}^j) = 100 \cdot \dfrac {l(\bm u_h(\mathcal{T}_{h,i}^j)) - l(\bm u_h(\mathcal{T}_{h,1}^j))}{l(\bm u_h(\mathcal{T}_{h,1}^j))}, \quad i = 2,3, \quad j = c, f,
	\end{equation}
	where $\bm u_h(\mathcal{T}_{h,i}^j)$ denotes the PolyDG displacement field approximation associated with the mesh $\mathcal{T}_{h,i}^j$, for $i = 1,2,3$ and $j = c, f$.\\
	It is to remark that, for a small enough softness parameter $\gamma$, the level set-fitted mesh is practically as accurate as the explicit L-shaped model in terms of compliance. Specifically, the relative error is below $1\%$ in the level set-fitted case, for both the meshes. On the contrary, the percentage error is greater in scenario ii), which tends to underestimate the compliance, in particular on the coarse tessellation where $\Delta_\% l (\mathcal{T}_{h,2}^c)$
	is twice as much bigger than $\Delta_\% l (\mathcal{T}_{h,3}^c)$. Moreover, the grid in ii) attains a similar accuracy level (i.e., $\simeq -0.85\%$) as scenario iii), only if appropriately refined, as highlighted by cross-comparing the third and the fourth column. Finally, on analyzing columns $4$ and $5$, we notice that the increased mesh resolution limits the differences in the percentage error.
	\begin{table}[H]
		\centering
		\begin{tabular}{c | c | c | c | c} 
			\hline
			$\gamma$ &
			$\Delta_\% l (\mathcal{T}_{h,2}^c)$ &
			$\Delta_\% l (\mathcal{T}_{h,3}^c)$ &
			$\Delta_\% l (\mathcal{T}_{h,2}^f)$ &
			$\Delta_\% l (\mathcal{T}_{h,3}^f)$ \\
			\hline
			$10^{-2}$  & $-35.774$   & $-34.561$   & $-35.750$  & $-35.056$ \\
			$10^{-4}$  & $-2.0611$   & $-1.3877$   & $-1.3882$  & $-1.1702$  \\
			$10^{-6}$  & $-1.5329$   & $-0.8768$   & $-0.8446$  & $-0.6424$  \\
			$10^{-8}$  & $-1.5276$   & $-0.8716$   & $-0.8391$  & $-0.6371$  \\
			$10^{-12}$ & $-1.5276$   & $-0.8716$   & $-0.8391$  & $-0.6370$  \\
			\hline
		\end{tabular}
		\caption{Test case of Section~\ref{sec:lshape_validation} - Percentage error on the compliance for different values of $\gamma$ and polygonal meshes, according to definition \eqref{eq:delta}.}\label{tab:compliance}
	\end{table}
	

	\section{Body-fitted meshes in topology optimization}
	\label{sec:topopt}
	
	The results in Section~\ref{sec:lshape_validation} motivate the employment of level set-fitted polygonal meshes in the presence of moving interfaces between material discontinuities. Among the diverse possible applications, we challenge the mesh fitting methodology in the context of structural topology optimization (TO). The rationale is twofold, consistently with the literature related to body-fitted methods for TO (\cite{li2022,zhuang2022,andreasen2020,yamasaki2011}): we limit the inaccuracy in the structural evaluation typical of a non-fitted approach \cite{nardoni2022}; we enhance the geometrical description of the structure under optimization thanks to the automatic insertion of mesh edges tracking the structure boundary. 
	However, with respect to traditional body-fitted approaches in TO, we propose a new PolyDG-based workflow, which allows us to eliminate computationally demanding local and global post-processing remeshing operations, thus leading to an effective and automatic structure design procedure.
	
	\subsection{The minimum compliance topology optimization}
	\label{sec:min_compliance}
	
	Topology optimization is a class of mathematical techniques that aim at devising the optimal distribution of material within a given design domain, while satisfying specific performance criteria~\cite{rozvany,sigmund2013,sigmund2004}. Through the last decades, TO proved to be extremely versatile in terms of applications, ranging from the standard structural design optimization framework to more complex physical settings, involving, for instance, multi-scale and multi-physics phenomena~\cite{alaimo2017,collet2018,Ferro2021,huang2012,Gavazzoni2022,carbonaro2023}. On top of the physical setting under consideration, a huge variety of design requirements have been investigated to guarantee the feasibility with respect to given prescriptions, e.g., volume occupation, manufacturing and geometric constraints~\cite{liu2018,ibhadode2023}.
	\\
	The mathematical formalization of a TO problem requires to describe the design under optimization and to account for the topology changes occurring throughout the process. With this aim, various numerical methods have been developed, including the widespread density-based approaches~\cite{sigmund2004}, the level set~\cite{allaire,yulin2004} and the homogenization methods~\cite{allaire2004,bendsoekikuchi1988}.
	Here, we focus on a level set approach, consistently with the contents in Section~\ref{sec:fittedmesh}.
	
	The level set method identifies the boundary of the structure under optimization as the zero-isoline of a signed distance function  and evolves such boundary in order to comply with the imposed optimization criteria. Specifically, we consider the signed-distance-like function $\varphi: \Omega \rightarrow [-1, 1]$ that partitions the design domain into the interior ($\Sigma$), the boundary ($\partial\Sigma$) and the exterior ($\cancel{\Sigma}$) portions of the structure, given by
		$\Sigma =   \{ {\bm x} \in \Omega : 0 < \varphi \le 1 \}$, $\partial\Sigma =\{ {\bm x} \in \Omega : \varphi({\bm x}) = 0 \}$ and $\cancel{\Sigma}  =  \{ {\bm x} \in \Omega : -1 \le \varphi < 0 \}$.
	Function $\varphi$, together with portions $\Sigma$, $\partial \Sigma$, $\cancel{\Sigma}$ are consistent with the definitions provided in Section~\ref{sec:fittedmesh}, here allowing $\varphi$ to take value in $[-1,1]$ only, in order to mimic a phase-field formulation, as proposed in~\cite{yamada2010topology}. The structure $\partial\Sigma \cup \Sigma$ is optimized by suitably changing the level set function in $\Omega$. These modifications are typically guided by an evolution process, governed by a partial differential equation, that tracks the movement of function $\varphi$ over a fictitious time domain and enables the systematic exploration of various material distributions and topologies~\cite{yamada2010topology,yaji2016shape}.
	
	In the context of structural optimization problems, TO proved to be effective in designing robust and lightweight configurations (see, e.g.,~\cite{ferro2020c}). With this regard, the minimum compliance problem is a simple yet elegant topology optimization formulation for redistributing or removing material inside the considered design domain, in order to devise stiff structures at a minimum material usage~\cite{sigmund2004,sigmund2013}. 
	The level set method resorts to the identification of material and void portions through variable $\chi_\varphi$, namely the characteristic function associated with $\varphi$, so that
	$$
	\chi_\varphi = \left\{
	\begin{array}{ll}
		1 & \varphi \ge 0, \\
		0 & \varphi < 0
	\end{array}
	\right.
	\iff
	\chi_\varphi = \left\{
	\begin{array}{ll}
		1 & \bm x \in \Sigma \cup \partial \Sigma, \\
		0 & \bm x \in \cancel{\Sigma}.
	\end{array}
	\right.
	$$
	The minimum compliance problem can be formulated as:
	{\it find $\varphi \in H^1(\Omega)$ such that}
	\begin{equation}\label{eq:minlevel_set}
		\min_{\varphi}\,l({\bm u}(\varphi)) \ :
		\left\{
		\begin{array}{l}
			\mathcal{A}^e(\bm u(\varphi), \bm v) = (\bm g, \bm v)_{\Gamma_N} \quad  \forall\, \bm v \in  [H^{1}_{\Gamma_D}(\Omega)]^2,\\[3mm]
			\displaystyle \displaystyle \int_\Omega \chi_\varphi d\bm x \le \alpha 
			{\rm Vol}_0,
		\end{array}
		\right.
	\end{equation}
	where $l$ is the compliance of the structure introduced in \eqref{eq:compliance} and here depending implicitly on $\varphi$ via $\bm u(\varphi)$; $\mathcal{A}^e(\bm u(\varphi), \bm v)$ is the bilinear form in \eqref{eq:linear_weak}, modelling the linear elasticity system, where the stiffness tensor $\mathbb D$ is $\mathbb D^0 \chi_\varphi+\gamma\mathbb D^0(1-\chi_\varphi)$ to distinguish the properties of solid and void~\cite{allaire}; $\alpha \in (0, 1)$ is used to impose a maximum allowable volume fraction with respect to the measure, ${\rm Vol}_0 = |\Omega|$, of the whole design domain.
	
	Following~\cite{yamada2010topology,ceze15,otomori2014}, in the spirit of a pseudo-time-dependent continuation process, problem \eqref{eq:minlevel_set} can be recast as the the following evolution process for $\varphi = \varphi(\bm x, t)$
	\begin{equation}
		\begin{cases}
			{\displaystyle \frac{\partial\varphi}{\partial t}}=\Upsilon \, d_{t}\overline{F}+\tau\Delta\varphi & \textrm{in } \Omega,\,\,\, t>0\\[3mm]
			{\displaystyle \frac{\partial\varphi}{\partial n}=0} & \textrm{on }\partial \Omega,\,\,\, t>0\\[3mm]
			{\displaystyle \varphi=\varphi^{0}} & \textrm{in } \Omega,\,\,\, t = 0.
		\end{cases}\label{evolution}
	\end{equation}
	In particular, $\tau > 0$ is a Tikhonov-like regularization parameter that tunes the smoothness of the level set and can regulate the geometric complexity of the structure, acting as a perimeter control~\cite{Leoni}. The term $d_{t}\overline{F}$ denotes the topological derivative of the Lagrangian functional 
	$$
	\overline{F} = \overline{F}({\bm u}, {\bm w}, \theta; \varphi) = l({\bm u}) + \big[l({\bm w})-\mathcal{A}^e(\bm u, \bm w)\big] +
	\theta\bigg(\int_\Omega \chi_\varphi \, d\Omega - \alpha {\rm Vol}_0\bigg),
	$$
	associated with the compliance, where $\bm u$ is the solution to \eqref{eq:minlevel_set}$_1$, and the Lagrangian multipliers ${\bm w} \in [H^{1}_{\Gamma_D}(\Omega)]^2$ and $
	\theta \in \mathbb{R}^+$ enforce the state equation and the volume constraint, respectively. More specifically, with reference to, e.g., ~\cite{yamada2010topology,sokolowski2009topological,novotny2003,cortellessa2023}, $d_{t}\overline{F}$ is computed as
	$$
	d_{t}\overline{F} = d_{t}{F} -\theta, \qquad d_t F = \overline{\bm \sigma}({\bm u}) : \bm \epsilon({\bm w}),
	$$
	with $d_{t}{F}$ the topological derivative contribution ascribed to the compliance only, and
	where $\overline{\bm \sigma}({\bm u}) = 2 \overline{\mu}\bm\epsilon({\bm u})+\overline{\lambda} \nabla \cdot \bm u I$ is the modified stress tensor, computed by employing the tweaked Young modulus and Poisson ratio
	$$
	\overline{E} = \dfrac{4 A_2^2}{A_1 + 2A_2}, \quad \overline{\nu} = \dfrac{A_1}{A_1 + 2A_2},
	$$
	being
	$$
	A_1 = -\dfrac{ 3 E (1 - \nu) (1 - 14 \nu + 15 \nu^2)}{2(1 + \nu)(7 - 5 \nu) (1 - 2\nu)^2}, \quad A_2 = \dfrac{15 E (1-\nu)}{2 (1 + \nu) (7 - 5\nu)}.
	$$
	The Lagrange multipliers, $\bm w$ and $\theta$, are given by
	\begin{equation}\label{theta}
		\bm w = \bm u, \quad \theta = \dfrac{\int_\Omega d_t F \ d\Omega}{{\rm Vol}_0} \exp \bigg( p_1  \dfrac{\int_\Omega \chi_\varphi(\varphi) \, d\Omega - \alpha {\rm Vol}_0}{\alpha {\rm Vol}_0} + p_1 p_2 \bigg),
	\end{equation}
	with $p_1$ and $p_2$ parameters to be tuned in order to enforce the volume constraint~\cite{yamada2010topology,otomori2014}.
	Finally, $\Upsilon > 0$ is a normalization factor chosen as
	\begin{equation}\label{upsilon}
		\Upsilon = \dfrac{{\rm Vol}_0}{\displaystyle\int_\Omega |d_t \overline{F}| \ d\Omega}.
	\end{equation}
	Problem \eqref{evolution} is completed with homogeneous Neumann boundary conditions for simplicity of implementation, and with an initial condition, $\varphi^0$, for $\varphi$ at $t = 0$.
	
	For a detailed description of the method and the involved parameters, we refer the interested reader to~\cite{yamada2010topology}.
	
	
	

	\subsection{The PolyDG discretization for the minimum compliance problem}
	\label{sec:poly_discretization}
	The discretization of the continuous level set problem~\eqref{evolution} with a PolyDG approach resorts to the functional space setting introduced in Sections~\ref{sec:fittedmesh_method}-\ref{sec:linearelasticity}. 
	According to a Symmetric Interior Penalty DG discretization~\cite{Riviere,Ern2021,Arnold82,Arnoldbrezzicockburnmarini2002,Wheeler78}, the semi-discrete parabolic problem in weak form reads as: {\it find $\varphi_h = \varphi_h(t) \in V_h$ for all $t > 0$, such that}
	\begin{equation}
		\begin{cases}
			\left( {\displaystyle \frac{\partial\varphi_h}{\partial t}}, \psi_h \right)_{\mathcal{T}_h} + \mathcal{B}(\varphi_h, \psi_h)
			=\left( \Upsilon \, d_{t}\overline{F}, \psi_h \right)_{\mathcal{T}_h} & \forall \psi_h \in V_h,\,\,\, t>0\\[5mm]
			{\displaystyle \varphi_h(0)= \Pi^{V_h} \varphi^{0}} & \textrm{in } \Omega,
		\end{cases}\label{evolution_discrete}
	\end{equation}
	where $\Pi^{V_h}$ is an $L^2(\Omega)$-projection operator onto space $V_h$, and the operator $\mathcal{B}(\varphi_h, \psi_h)$ is defined as
	$$
	\begin{array}{lcl}
		\mathcal{B}(\varphi_h, \psi_h) &=& (\tau\nabla_h\varphi_h, \nabla_h\psi_h)_{\mathcal{T}_h}
		-( \llbracket \tau \nabla_h \varphi_h  \rrbracket, \llbrace \psi_h \rrbrace)_{\mathcal{F}_h^i\cup\mathcal{F}_h^D}
		\\[3mm]
		&-&( \llbracket \varphi_h  \rrbracket, \llbrace \tau \nabla_h \psi_h \rrbrace)_{\mathcal{F}_h^i\cup\mathcal{F}_h^D}
		+  (\eta_\tau\,\llbracket \varphi_h  \rrbracket, \llbracket \psi_h  \rrbracket)_{\mathcal{F}_h^i\cup\mathcal{F}_h^D},
	\end{array}
	$$
	with $\eta_\tau:\mathcal{F}_h\rightarrow\mathbb{R}^+$ the penalization factor given by
	\begin{equation*}
		\eta_\tau=\sigma_{0, \tau} \, \tau
		\begin{cases}
			\underset{\kappa\in\{\kappa_1,\kappa_2\} } \max  C(p, \kappa, F), & F \in \mathcal{F}_h^i, \, F \subseteq \partial \kappa_1 \cap \partial \kappa_2, \\
			C(p, \kappa, F), &  F\in\mathcal{F}_h^D,  \, F \subset \partial \kappa \cap \Gamma_D,
		\end{cases}
	\end{equation*}
	and $\sigma_{0,\tau}$ a (large enough) positive constant value to be assigned.
	
	The fully-discrete evolution problem on polytopal meshes requires solving equation \eqref{evolution_discrete} on a partition of the time window under consideration of uniform spacing $\Delta t$, namely by introducing the discrete time instants $\{t^{\tt k}\}_{{\tt k = 0}}^{{\tt k}_{\rm max}}$, with $t^{\tt 0} = 0$ and  ${\tt k}_{\rm max}$ the maximum number of time steps and $t^{\tt k + 1} - t^{\tt k} = \Delta t$, for $\tt k \ge 0$.
	By resorting to a semi-implicit scheme, we formalize the evolution of the level set function in a TO problem as: {\it{for ${\tt k} =  {\tt 0}, ..., {\tt k}_{\rm max}- {\tt 1}$, find $\varphi_h^{{\tt k }+1} \in V_h$, such that}}
	\begin{equation}\label{time_adv}
		\left(\dfrac{\varphi_h^{{\tt k + 1}} - \varphi_h^{{\tt k}}}{\Delta t} , \psi_h \right)_{\mathcal{T}_h}
		+ \mathcal{B}(\varphi_h^{\tt k + 1}, \psi_h)= (\Upsilon d_{t} \overline{F}^{{\tt k}}, \psi_h )_{\mathcal{T}_h} \quad \forall \psi_h \in V_h, {\tt k } \ge {\tt 0},
	\end{equation}
	where the contribution of the topological derivative is explicitly evaluated at $t^{\tt k}$.
	
	\subsection{The topology optimization algorithm with level set-fitting }
	\label{sec:fitting_algorithm}
	The combination of the fitting procedure associated with the level set variable $\varphi$ in Section~\ref{sec:fittedmesh} and the topology optimization scheme in \eqref{time_adv} leads to a new topology optimization pipeline with respect to what available in the literature~\cite{otomori2014}.
	
	The proposed strategy is based on two computational meshes, i.e., a fixed grid $\mathcal{T}_h^{\rm \, ls}$ for the level set evolution and a level set-fitted mesh $\mathcal{T}_h^{\rm el}$ for the linear elasticity equation approximation. This choice is already employed in other works (see, e.g.,~\cite{furuta2023}) and allows us to evolve the level-set onto a shape-regular tessellation, thus avoiding any bias on the optimized layout due to the introduction of new mesh edges by the fitting procedure.
	We indicate by the superscripts ${\rm ls}$ and ${\rm el}$ the variable associated with the first or the second grid, respectively, only in case of ambiguity.\\
	The pseudocode is listed in Algorithm~\ref{algorithm}.
	
	The topology optimization procedure consists of a {\bf while} loop, whose internal instructions are repeated until a termination condition on the compliance stagnation or on the maximum number of iterations is met. At line \ref{algo:project1}, the level set $\varphi_h^{\tt k, \rm ls}$ is projected onto the grid $\mathcal{T}_h^{\rm el}$ by function {\tt project}, which implements a standard $L^2$-projection, with a view to the PolyDG linear elasticity solving. 
	Then, function {\tt solveState} returns the solution to equation \eqref{eq:linear_discrete}, for $\mathcal{T}_h = \mathcal{T}_h^{\rm el}$ and where the material parameters are assigned through the discrete indicator function, $\chi_{\varphi,h}^{\tt k, \rm el}$, and the softness parameter $\gamma$ (line \ref{algo:solve}). 
	Line \ref{algo:derivative} assembles the topological derivative terms by employing the finite element functions defined on $\mathcal{T}_h^{\rm el}$, as detailed in Section~\ref{sec:min_compliance}. Successively, at line~\ref{algo:project2}, quantity $d_{t} F^{\tt k, \rm el}$ is projected onto the level set grid through function {\tt project}. 
	The evolution of the level set is carried out by function {\tt evolveLevelSet}, which implements \eqref{evolution_discrete} on $\mathcal{T}_h = \mathcal{T}_h^{\rm ls}$ and returns the updated level set function $\tilde{\varphi}_h^{\tt k+1, \rm ls}$, which, in principle, is not guaranteed to be bounded in the interval $[-1, 1]$ (line~\ref{algo:evolve}). For this reason, function {\tt threshold} in line \ref{algo:thresh} applies a cutoff operation, so that $ -1 \le {\varphi}_h^{\tt k+1, \rm ls} \le 1$.
	\\
	As a next step, we generate a mesh for the displacement computation, which is fitted to the current design variable. With this aim, we resort to the procedure {\tt fitMesh} detailed in Algorithm~\ref{cutalgorithm}. 
	However, when dealing with discontinuous finite elements, the routine in Section~\ref{sec:fittedmesh_method} might yield local zero-curves, $\ell_h^\kappa$, that are not continuously connected across the elements, thus identifying a potentially discontinuous structural contour, $\mathcal{C}_h$.
	Although this is a viable option for a generic PolyDG method, it does not offer an ideal setting for TO applications, where a continuous interface is expected. To retrieve a continuous form for the contour $\mathcal{C}_h$, we feed the {\tt fitMesh} procedure with the variable $\Pi^C \varphi_h^{\tt k+1, \rm ls}$, namely the projection of $\varphi_h^{\tt k+1, \rm ls}$ onto a suitable continuous space (line~\ref{algo:fit}). We highlight that Algorithm~\ref{cutalgorithm} is invoked every {\tt kCut} iterations, starting from ${\tt k} \ge {\tt kStart}$, thus allowing the user to tune the computational overhead introduced by the fitting procedure.\\
	Finally, we compute the relative error on the compliance and we assume that the convergence is attained if errComp is to within the prescribed tolerance for two successive iterations (line \ref{algo:error}).
	
	\begin{algorithm}[H]
		\caption{Minimum compliance topology optimization with level set-fitted polygonal meshes}\label{algorithm}
		{\bf Input} :  {\tt CTOL, kmax, kStart, kCut}, $\varphi^{\tt 0}$, $\mathcal{T}_h^{\rm \, ls}$, $\Delta t$, $\alpha$, $\gamma$, $\tau$
		\begin{algorithmic}[1]
			\State Set: $\mathcal{T}_h^{\rm el} = \mathcal{T}_h^{\rm \, ls}$, $\varphi_h^{\tt 0, \rm ls}=\Pi^{V_h}\varphi^{\tt 0}$, ${\tt k} = {\tt 0}$, errComp = 1+{\tt CTOL}
			\vspace{1mm}
			\While { errComp $> {\tt CTOL}$ \& ${\tt k} <$ {\tt kmax}}
			\vspace{2mm}
			\State [$\varphi_h^{\tt k, \rm el}$, $\chi_{\varphi,h}^{\tt k, \rm el}$] = {\tt project}($\varphi_h^{\tt k, \rm ls}$,            $\mathcal{T}_h^{\rm el}$);
			\label{algo:project1} \vspace{2mm}
			\State ${\bm u}_h^{\tt k}$ = {\tt solveState}($\mathcal{T}_h^{\rm el}$, $\chi_{\varphi,h}^{\tt k, \rm el}$, $\gamma$);
			\label{algo:solve} \vspace{2mm}
			\State [$d_{t} F^{\tt k, \rm el}$, $\theta^{\tt k}$, $\Upsilon$] = {\tt topologicalDerivative}(${\bm u}_h^{\tt k}$, $\varphi_h^{\tt k, \rm el}$, $\alpha$, ${\rm Vol}_0$);
			\label{algo:derivative}\vspace{2mm}
			\State [$d_{t} F^{\tt k, \rm ls}$] = {\tt project}($d_{t} F^{\tt k, \rm el}$, $\mathcal{T}_h^{\rm \, ls}$);
			\label{algo:project2}\vspace{2mm}
			\State $\tilde{\varphi}_h^{\tt k+1, \rm ls}$ = {\tt evolveLevelSet}($\mathcal{T}_h^{\rm \, ls}$, $\varphi_h^{\tt k, \rm ls}$, $d_{t} F^{\tt k, \rm ls}$, $\theta^{\tt k}$, $\Upsilon$, $\tau$, $\Delta t$);
			\label{algo:evolve}\vspace{2mm}
			\State $\varphi_h^{\tt k+1, \rm ls}$ = {\tt threshold}($\tilde{\varphi}_h^{\tt k+1, \rm ls}$);
			\label{algo:thresh}\vspace{2mm}
			\If {${\tt k}\ge{\tt kStart}$ \& ${\tt mod}({\tt k},{\tt kCut})==0$}
			\label{algo:if}\vspace{2mm}
			\State $\mathcal{T}_h^{\rm el}$ = {\tt fitMesh}($\mathcal{T}_h^{\rm \, ls}$, $\Pi^C \varphi_h^{\tt k+1, \rm ls}$);
			\label{algo:fit}\vspace{2mm}
			\EndIf
			\State {\bf end if}
			\label{algo:endif}\vspace{2mm}
			\If  {${\tt k} \ge 2$}
			\vspace{2mm}
			\State errComp$^1$ = {\tt computeError}(${\bm u}_h^{\tt k}$, ${\bm u}_h^{\tt k-1}$);
			\vspace{2mm}
			\State errComp$^2$ = {\tt computeError}(${\bm u}_h^{\tt k-1}$, ${\bm u}_h^{\tt k-2}$);
			\vspace{2mm}
			\State errComp = max(errComp$^1$, errComp$^2$);
			\label{algo:error}\vspace{2mm}
			\EndIf
			\State {\bf end if}
			\vspace{2mm}
			\State ${\tt k} = {\tt k+1}$;
			\vspace{2mm}
			\EndWhile
			\State {\bf end while}
			\vspace{2mm}
			\State [$\varphi_h^{\tt k, \rm el}$, $\chi_{\varphi,h}^{\tt k, \rm el}$] = {\tt project}($\varphi_h^{\tt k, \rm ls}$, $\mathcal{T}_h^{\rm el}$);
			\vspace{2mm}
		\end{algorithmic}
		{\bf Output}: $\varphi_h^{\tt k, \rm el}$, $\chi_{\varphi,h}^{\tt k, \rm el}$, $\mathcal{T}_h^{\tt k, \rm el}$
	\end{algorithm}
	
	\noindent
	Algorithm~\ref{algorithm} returns the optimized level set function $\varphi_h^{\tt k, \rm el}$, the corresponding indicator function $\chi_{\varphi,h}^{\tt k, \rm el}$, and the fitted computational grid $\mathcal{T}_h^{\tt k, \rm el}$.
	
	Few comments are in order. 
	Concerning the projection step at lines~\ref{algo:project1} and~\ref{algo:project2}, we remark that no projection error occurs when mapping $\varphi_h^{\tt k, \rm ls}$ into
	$\varphi_h^{\tt k, \rm el}$, since the edges of $\mathcal{T}_h^{\rm \, ls}$ represent a subset of the edges of $\mathcal{T}_h^{\rm el}$. The same way of reasoning does not hold when projecting $d_{t} F^{\tt k, \rm el}$ in $d_{t} F^{\tt k, \rm ls}$, the projection taking place now from a finer to a coarser mesh.
	
	Moreover, a standard level set-based topology optimization algorithm without any fitting procedure can be retrieved from Algorithm~\ref{algorithm}, by setting ${\tt kStart} > {\tt kmax}$, i.e., by skipping any update operation for $\mathcal{T}_h^{\rm el}$. In such a case, the projections in lines \ref{algo:project1} and \ref{algo:project2} can be skipped too.
	
	\subsection{Numerical results}
	\label{sec:numerical}
	In this section, we test Algorithm~\ref{algorithm} on three test cases in order to investigate the good properties of the proposed method. We challenge the algorithm with the optimization of a standard cantilever beam configuration, of an L-shaped domain, and, finally, of a more challenging geometry. \\ The implementation of the proposed approach is carried out in {\tt Matlab}, where we resort to an in-house code for the PolyDG solver, to {\tt Polymesher}~\cite{talischi2012polymesher} for the creation of the input mesh $\mathcal{T}_h^{\rm \, ls}$ and to {\tt Polygon Clipper}~\cite{polygonclipper} for the mesh fitting operations.
	
	In the simulations below, some parameters are varied to assess the corresponding sensitivity of the methodology, while some values are kept fixed starting from reference values in the literature~\cite{Ern2021,Riviere,yamada2010topology,cortellessa2023}. In particular, in all the test cases, we set $\sigma_{0, \mu} = \sigma_{0, \lambda} = \sigma_{0, \tau} = 10$, $p_1 = 4$, $p_2 = -0.02$, {\tt CTOL}$ = 1$e$-4$, {\tt kmax}$ = 300$, $\gamma = 10^{-3}$. 
	
	\subsubsection{The cantilever beam}\label{sec:cantilever}
	As a first verification test, we consider a standard benchmark case, namely the topology optimization of a cantilever beam. The domain $\Omega = (0,80) \times (0, 50)$ is clamped at the left boundary, i.e., $\Gamma_D = \{\bm x \in \partial \Omega : x = 0\}$, and is loaded on $\Gamma_N = \{\bm x \in \partial \Omega : x = 80\} $ with a traction $\bm g = [0, -50 \ \mathbb{I}_{[24,26]}(y)]^T$. Inspired by the data used in~\cite{zhuang2021}, the parameters for the base material are $E^0 = 1e5$ and $\nu^0 = 0.3$.
	
	Algorithm~\ref{algorithm} is run by setting  {\tt kStart}$ = 0$, {{\tt kCut} = 5}, $\Delta t = 0.05$, $\alpha = 0.5$, $\tau = 1.5$, a computational mesh $\mathcal{T}_h^{\rm \, ls}$ consisting of $1000$ polygons, and a uniform material distribution with a hole of radius $10$ centered at $(40,25)$ as initial condition $\varphi^0$.
	
\begin{figure}[h!]
		\centering
		\includegraphics[width=0.975\textwidth, keepaspectratio]{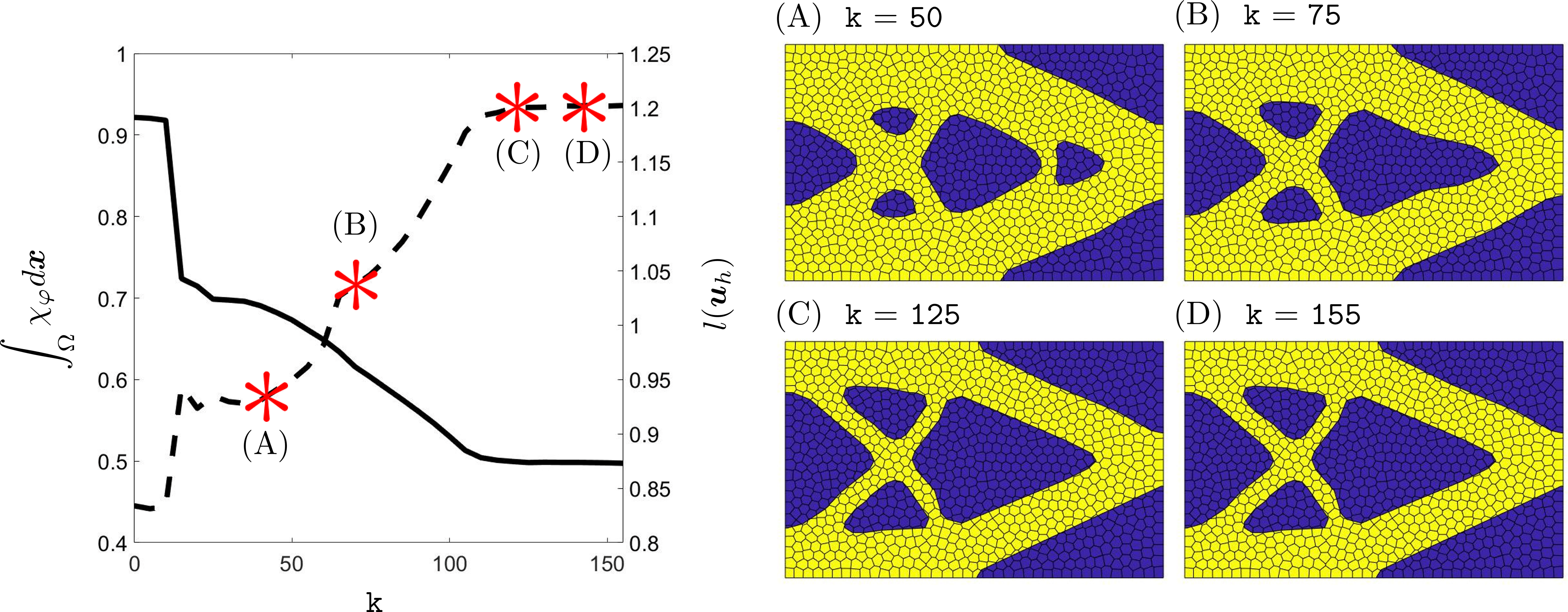} 
		\caption{Test case of Section~\ref{sec:cantilever} - Convergence history (left) of the compliance (dashed line) and of the volume fraction (solid line); indicator function $\chi_{\varphi, h}^{\tt k, \rm el}$ superimposed to the fitted mesh (right) at iterations ${\tt k} = {\tt 50}, {\tt 75}, {\tt 125}, {\tt 155}$, red-highlighted in the left panel.}
		\label{fig:cantilever_zhang_convergence}
	\end{figure}

	Figure~\ref{fig:cantilever_zhang_convergence} (left) presents the convergence history for the compliance functional $l$ and the volume constraint. We observe the expected trend, consisting in the reduction of the volume until the upper bound of the constraint is reached and the subsequent stagnation of the compliance.
	Figure~\ref{fig:cantilever_zhang_convergence} (right) shows the fitted mesh together with the indicator function at four time instants during the level set evolution, in order to appreciate how the fitted mesh $\mathcal{T}_h^{\rm el}$ perfectly catches the interface and separates the topology under optimization from the soft material (i.e., the void). Panel (D), basically coinciding with panel (C), displays the converged structure at iteration ${\tt k} = {\tt 155}$, which is characterized by a compliance equal to $1.2021$, a volume fraction equal to $49.74\%$, and by a topology that is recurrent in the literature (see, e.g.,~\cite{zhuang2021}).
	
\begin{figure}[h!]
		\centering
		\includegraphics[width=0.325\textwidth, keepaspectratio]{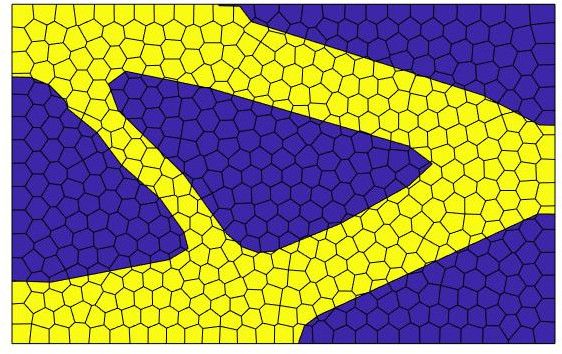}
		\includegraphics[width=0.325\textwidth, keepaspectratio]{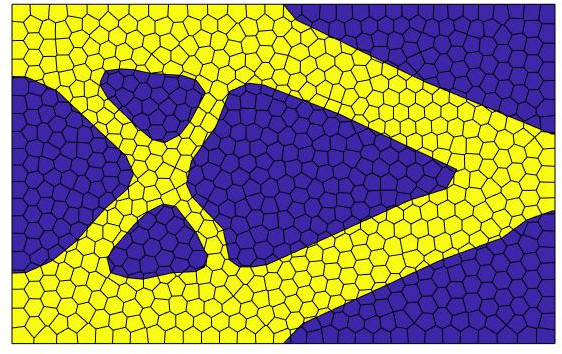}
		\includegraphics[width=0.325\textwidth, keepaspectratio]{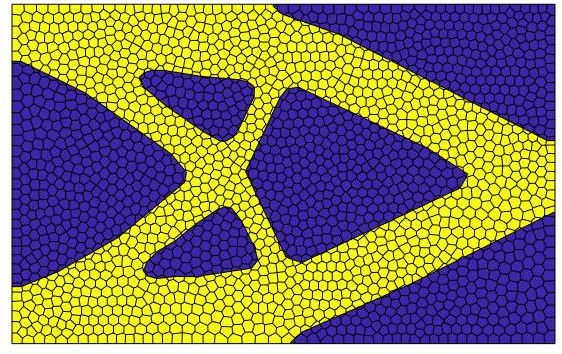} \\[3mm]
		\includegraphics[width=0.325\textwidth, keepaspectratio]{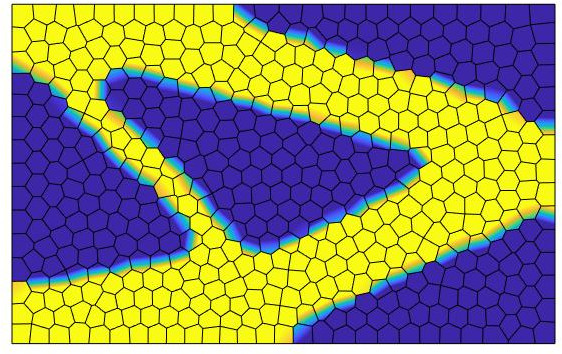}
		\includegraphics[width=0.325\textwidth, keepaspectratio]{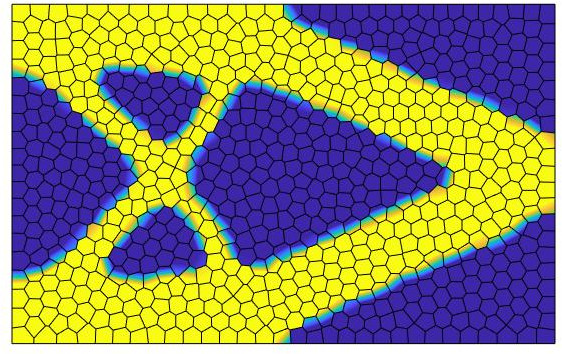}
		\includegraphics[width=0.325\textwidth, keepaspectratio]{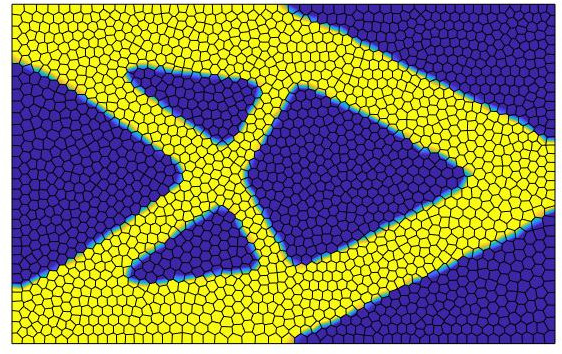}
		\caption{Test case of Section~\ref{sec:cantilever} - Indicator function $\chi_{\varphi, h}^{\tt k, \rm el}$ superimposed to the mesh for a level set-fitted (top) and a non-fitted (bottom) approach, for a different cardinality $\# \mathcal{T}_h^{\rm \, ls}$ of the level set grid (from left to right, $\# \mathcal{T}_h^{\rm \, ls} = 500, 750, 2000$).}
		\label{fig:cantilever_zhang}
	\end{figure}

	As expected, it can be checked that the discrepancy between the level-set fitted and the non-fitted approaches in terms of sharpness of the structure boundary is more striking when starting from a coarse mesh, with a view to a computationally cheap design procedure. This feature is confirmed in Figure~\ref{fig:cantilever_zhang} that showcases the optimized indicator function for the choice of $\mathcal{T}_h^{\rm \, ls}$ consisting of $500$, $750$ and $2000$ elements, when resorting to a level set-fitted (top row) or to a non-fitted (bottom row) approach. 
	We notice that the level set-fitted methodology is able to exactly match the level set zero-isoline and thus provides a crisp description of the configuration. Concerning the design yielded by the non-fitted approach,
	the representation of $\chi_{\varphi, h}^{\tt k, \rm el}$ through a function in $V_h$ leads to visualize the co-presence of solid material and void inside the same polygon, with a consequent irregular description of the structure boundary (see~\cite{zhuang2021} for similar issues).\\
	The results in the top panels are yielded by fitted meshes, $\mathcal{T}_h^{\rm el}$, consisting of $617$, $914$, and $2274$ polygonal elements for the three initial grids, $\mathcal{T}_h^{\rm ls}$. This proves that the superior geometrical accuracy is guaranteed with just a slight increase in terms of mesh cardinality, here limited to an increment of about $20\%$ elements.
	
	\subsubsection{The L-shaped configuration}\label{sec:lshape_TO}
	The topology optimization of an L-shaped domain is commonly investigated in the TO community (see, e.g.,~\cite{ferro2020b,nguyen2022}), since exhibiting some interesting aspects to tackle, especially in the stress distribution inside the structure. \\
	The physical setting is identified by $\Omega = (0,10)^2 \setminus [4, 10]^2$, with $\Gamma_D = \{\bm x \in \partial \Omega : y = 10\}$ and $\Gamma_N = \{\bm x \in \partial \Omega : x = 10\}$. The downward external traction is $\bm g = [0, -6 \ (1.5 - y)(y - 2.5) \ \mathbb{I}_{[1.5, 2.5]}(y) ]^T$ and the material parameters are $E^0 = 1$ and $\nu^0 = 0.3$.\\
	As far as Algorithm~\ref{algorithm} is concerned, we set {\tt kStart} $= 20$, {\tt kCut} $= 1$, $\Delta t = 5$e$-3$, $\alpha = 0.3$, $\tau = 2$, and $\varphi^0$ the function identically equal to $1$. 
	
\begin{figure}[h!]
		\centering
		\includegraphics[width=0.45\textwidth, keepaspectratio]{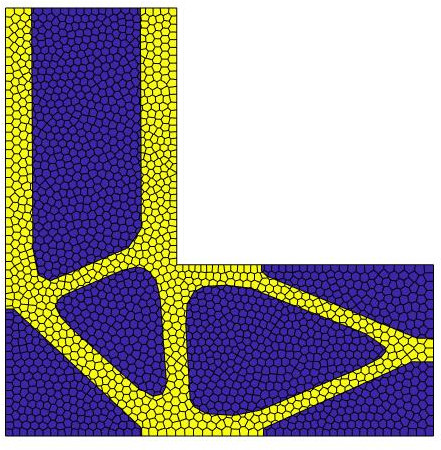} 
		\includegraphics[width=0.45\textwidth, keepaspectratio]{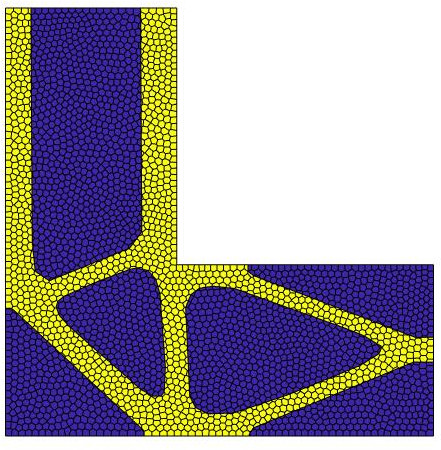} \\[3mm]
		\includegraphics[width=0.45\textwidth, keepaspectratio]{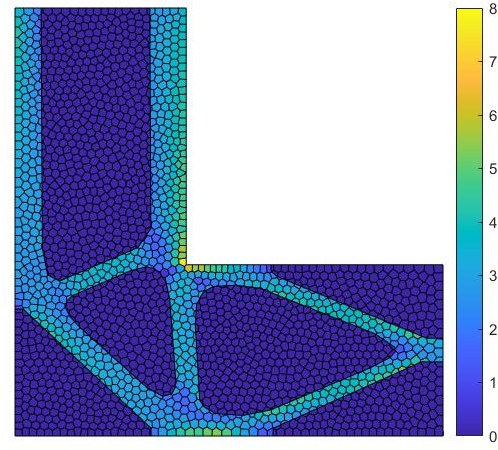} 
		\includegraphics[width=0.45\textwidth, keepaspectratio]{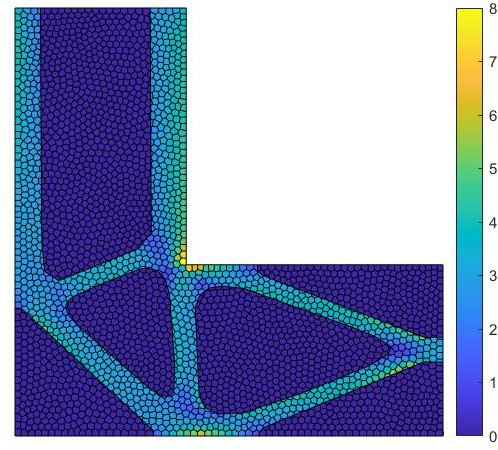}
		\caption{Test case of Section~\ref{sec:lshape_TO} - Indicator function $\chi_{\varphi, h}^{\tt k, \rm el}$ (top) and corresponding von Mises stress distribution (bottom)  superimposed to the fitted mesh, for a different cardinality $\#\mathcal{T}_h^{\rm \, ls}$ of the level set grid ($\#\mathcal{T}_h^{\rm \, ls}=2000$, left; $\#\mathcal{T}_h^{\rm \, ls}=3000$, right).}
		\label{fig:Lshape}
	\end{figure}

\begin{figure}[h!]
		\centering
		\includegraphics[height=0.4\textwidth, keepaspectratio]{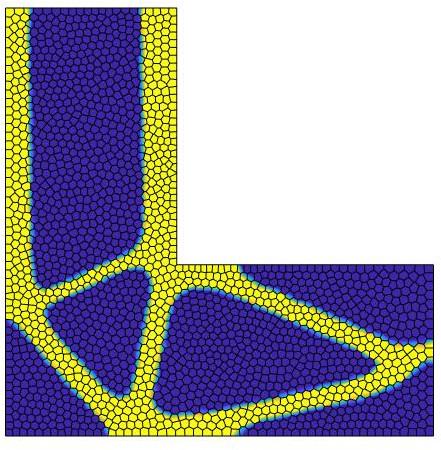} 
		\includegraphics[height=0.4\textwidth, keepaspectratio]{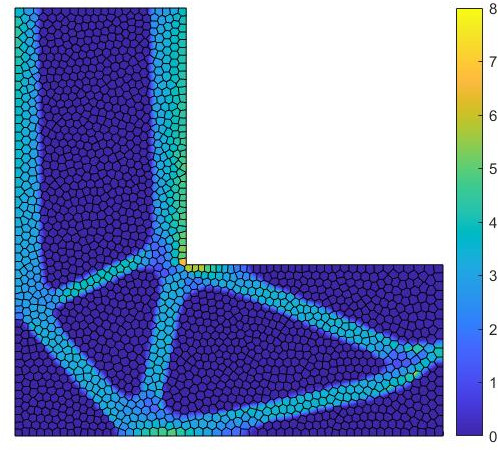} 
		\caption{Test case of Section~\ref{sec:lshape_TO} - Indicator function $\chi_{\varphi, h}^{\tt k, \rm el}$ (left) and corresponding von Mises stress distribution (right)  superimposed to the non-fitted mesh, for $\#\mathcal{T}_h^{\rm \, ls}=2000$.}
		\label{fig:compara}
	\end{figure}

	In Figure~\ref{fig:Lshape} (top), we show the output of the algorithm corresponding to two different starting meshes $\mathcal{T}_h^{\rm \, ls}$, characterized by $2000$ and $3000$ elements, respectively. 
	The final layouts, provided after $179$ and $198$ iterations, share the same topology and volume fraction equal to $30\%$, and are characterized by very similar compliance values ($219.3$ versus $222.3$ for the $2000$- and $3000$-polygon mesh, respectively), despite the different mesh resolution. This behavior can be ascribed to the use of the diffusivity parameter $\tau$, which effectively controls the geometrical complexity of the final layout~\cite{cortellessa2023,yamada2010topology}. 
	The final meshes $\mathcal{T}_h^{\rm el}$ consist of $2333$ and $3400$ elements, amounting to an added number of polygons lower than $15\%$ of the original cardinality. \\
	For comparison purposes, we run Algorithm~\ref{algorithm} on the coarse mesh by switching the fitting module off, while preserving all the other inputs. The left panel in Figure~\ref{fig:compara} shows the discrete indicator function $\chi_{\varphi, h}^{\tt k, \rm el}$ together with the associated non-fitted grid returned after $177$ iterations. We can drawn conclusions fully consistent with what observed about Figure~\ref{fig:cantilever_zhang} (bottom).\\
	The comparison between the level set-fitted and the non-fitted approaches becomes of interest from an application perspective when considering the distribution of the von Mises stress. Indeed, it is possible to appreciate the sharp localization of this field inside the structure in Figure~\ref{fig:Lshape}
	(bottom). On the contrary, the non-fitted discretization admits a diffused distribution of the von Mises stress along the structure interface (see Figure~\ref{fig:compara} (right)). This possibly leads to the storing of residual stresses in the weak material, which affects the local stress distribution along the boundaries, as remarked in~\cite{nardoni2022}.
	
\begin{figure}[h!]
		\centering	\includegraphics[width=0.75\textwidth, keepaspectratio]{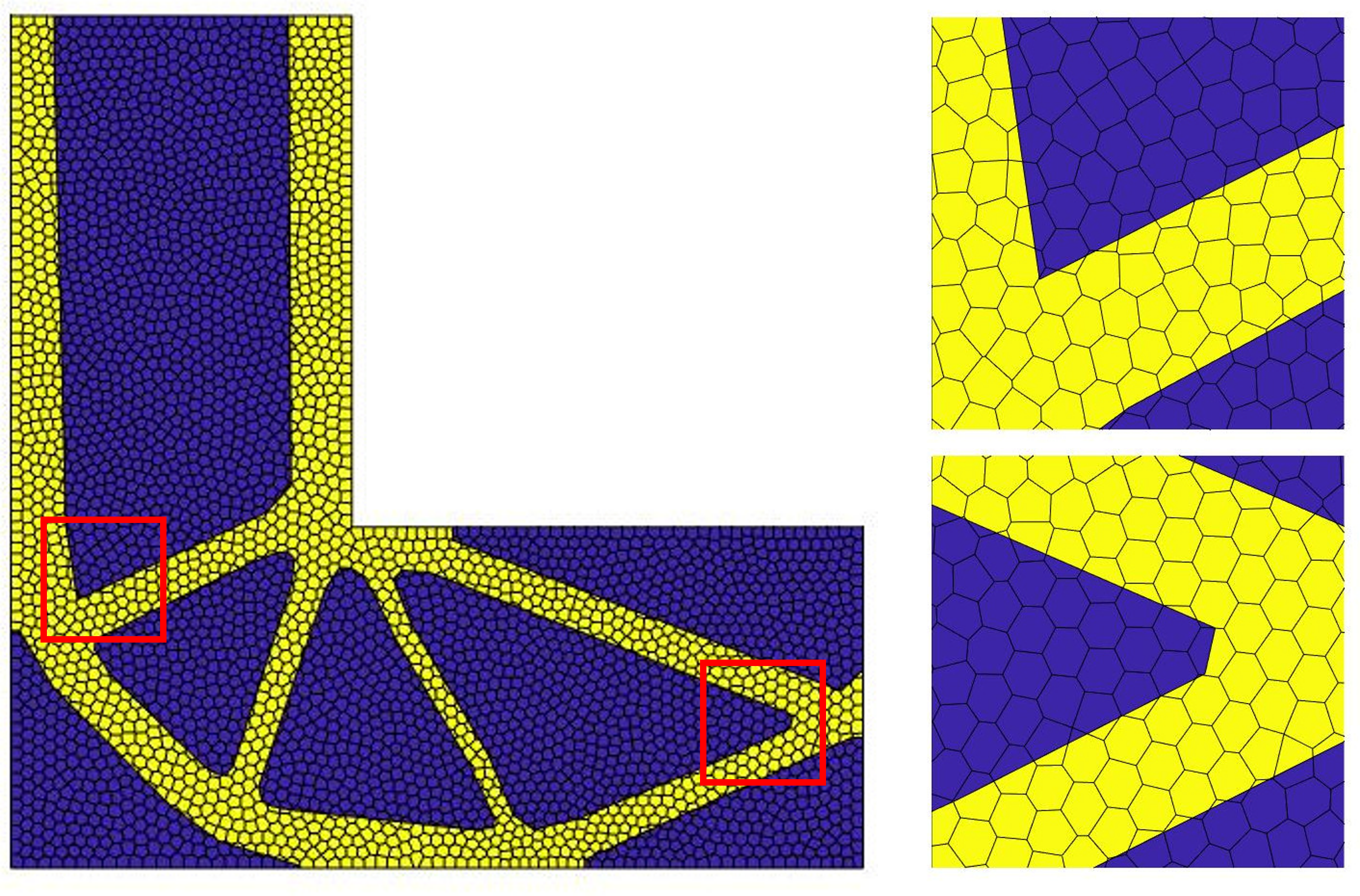} 
		\caption{Test case of Section~\ref{sec:lshape_TO} - Indicator function $\chi_{\varphi, h}^{\tt k, \rm el}$ superimposed to the level set-fitted mesh (left) and enlarged views (right), for $\#\mathcal{T}_h^{\rm \, ls}=3500$.}
		\label{fig:Lshape_zoom}
	\end{figure}

	This test case is also exploited to assess the sensitivity of the output layout with respect to the geometrical complexity parameter, $\tau$. In more detail, we expect that the greater the diffusivity, the simpler the achievable geometry. Figure~\ref{fig:Lshape_zoom} displays the output of Algorithm~\ref{algorithm} for $\tau = 0.25$ and a level set polygonal mesh $\#\mathcal{T}_h^{\rm \, ls}$ with $3500$ elements. The indicator function shows that the algorithm modifies the width and the number of thin struts stemming from the re-entrant corner when reducing $\tau$. Also in this case,   
	the mesh perfectly matches the boundary of the 
	internal structures thus highlighting the versatility of the level set-fitted approach.
	The final mesh $\mathcal{T}_h^{\rm el}$, which consists of $4014$ elements (corresponding to a $14.7\%$ increment of $\#\mathcal{T}_h^{\rm \, ls}$), exhibits elements of arbitrary shape and size, which do not compromise the robustness and the accuracy of the PolyDG solver~\cite{CangianiDongGeorgoulisHouston_2017}. \\
	The two enlarged views on the right show the peculiarly shaped polygons that can arise throughout the fitting procedure. Indeed, it is possible to spot some polytopal elements, that are divided almost equally by the edge insertion procedure, while others that become very thin and stretched.
	
	Finally, by qualitatively comparing  Figures~\ref{fig:Lshape} and \ref{fig:Lshape_zoom}, it is possible to notice that a large value for $\tau$ favors smooth geometries featuring rounded corners, while, on the contrary, small values of the diffusivity parameter leads to sharper corners (see, e.g., the top zoom in Figure~\ref{fig:Lshape_zoom}).
	
	\subsubsection{The wrench geometry}\label{sec:wrench}
\begin{figure}[hbt]
		\centering
		\includegraphics[width=0.75\textwidth, keepaspectratio]{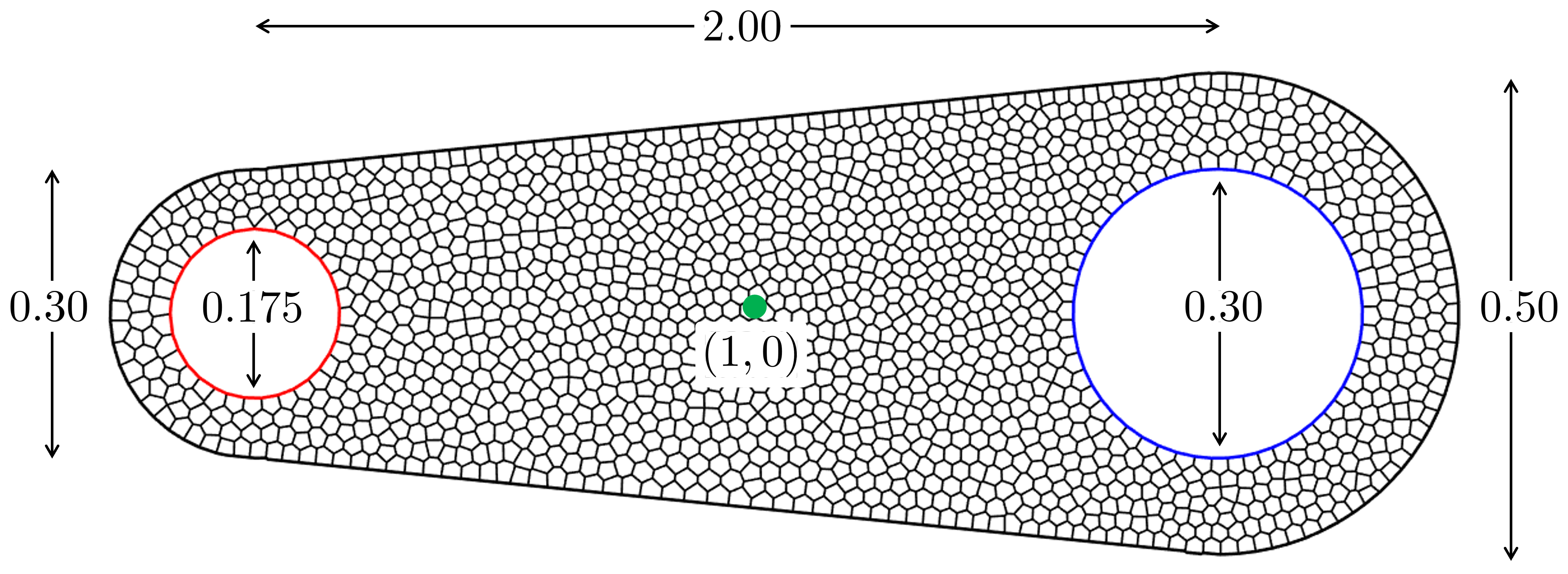} 
		\caption{Test case of Section~\ref{sec:wrench} - Geometry and polygonal computational mesh with non-homogeneous Neumann (red-highlighted) and homogeneous Dirichlet (blue-highlighted) boundary portions.}
		\label{fig:wrench_mesh}
	\end{figure}

\begin{figure}[h!]
		\centering
		\includegraphics[width=0.475\textwidth, keepaspectratio]{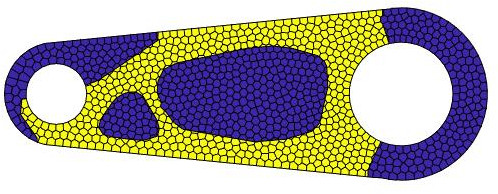}
		\includegraphics[width=0.475\textwidth, keepaspectratio]{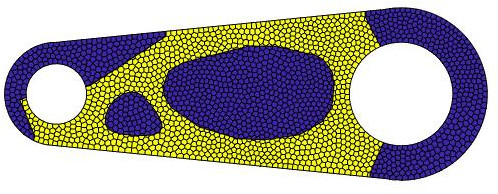}
		\\
		\includegraphics[width=0.475\textwidth, keepaspectratio]{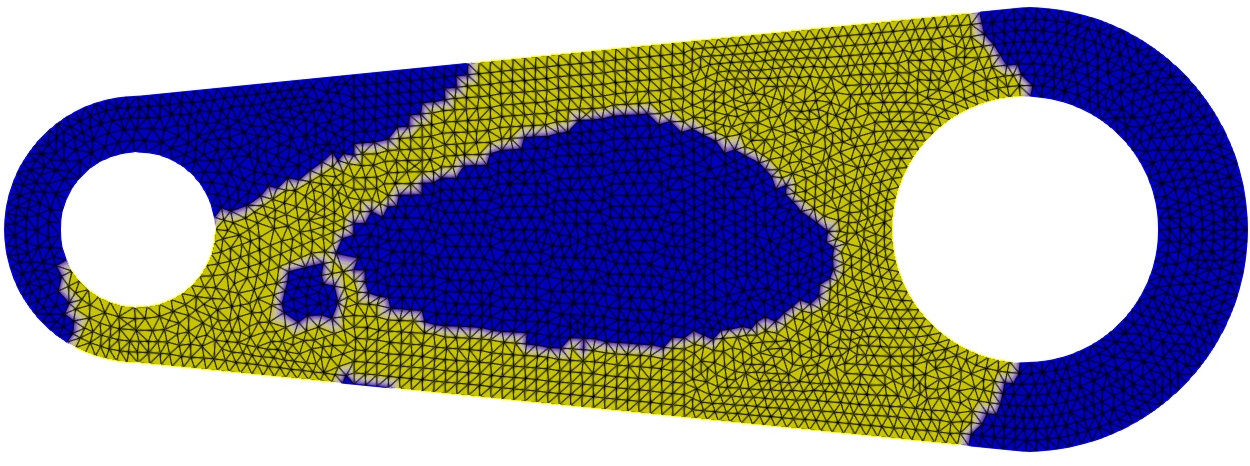}
		\includegraphics[width=0.475\textwidth, keepaspectratio]{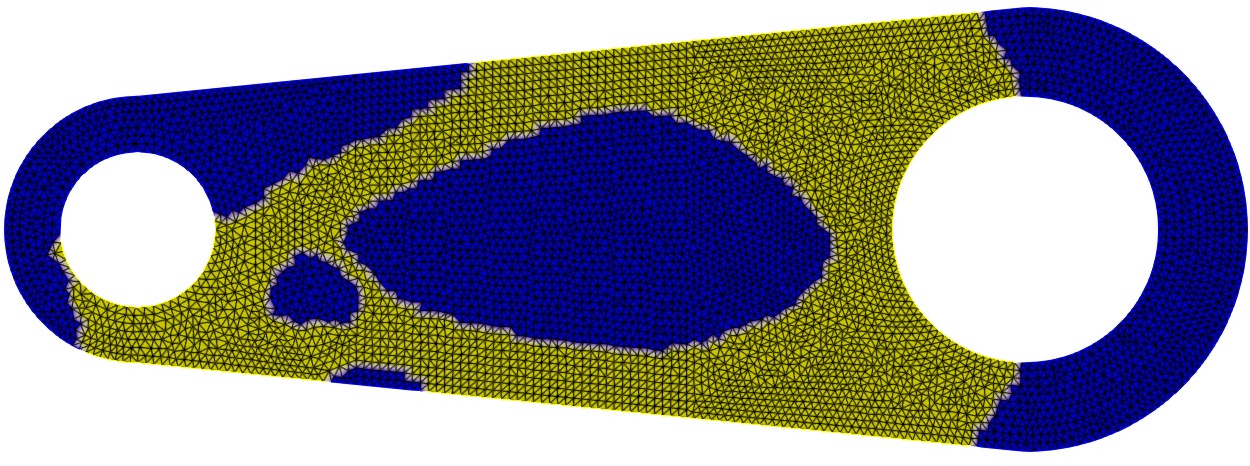}
		\\
		\includegraphics[width=0.475\textwidth, keepaspectratio]{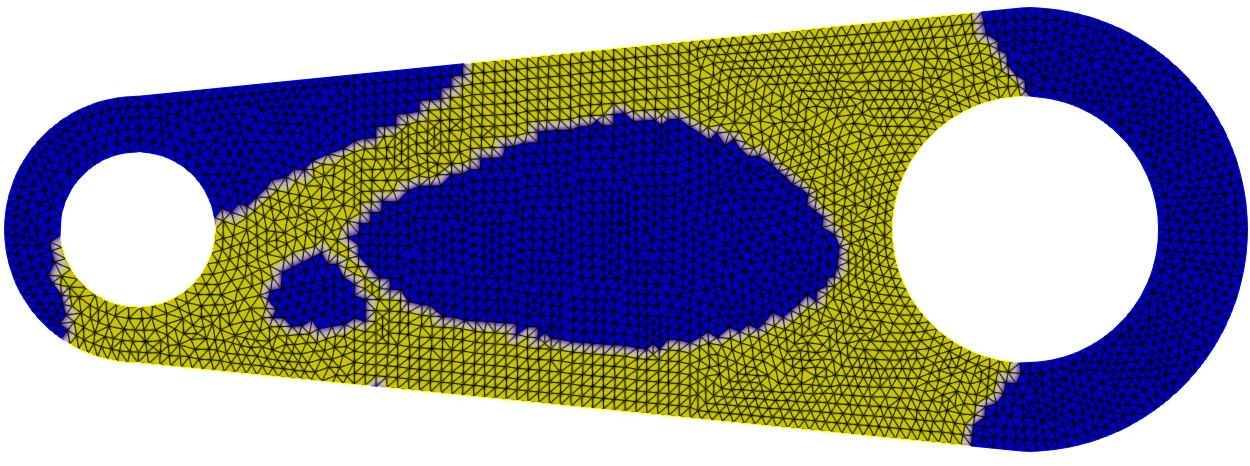}
		\includegraphics[width=0.475\textwidth, keepaspectratio]{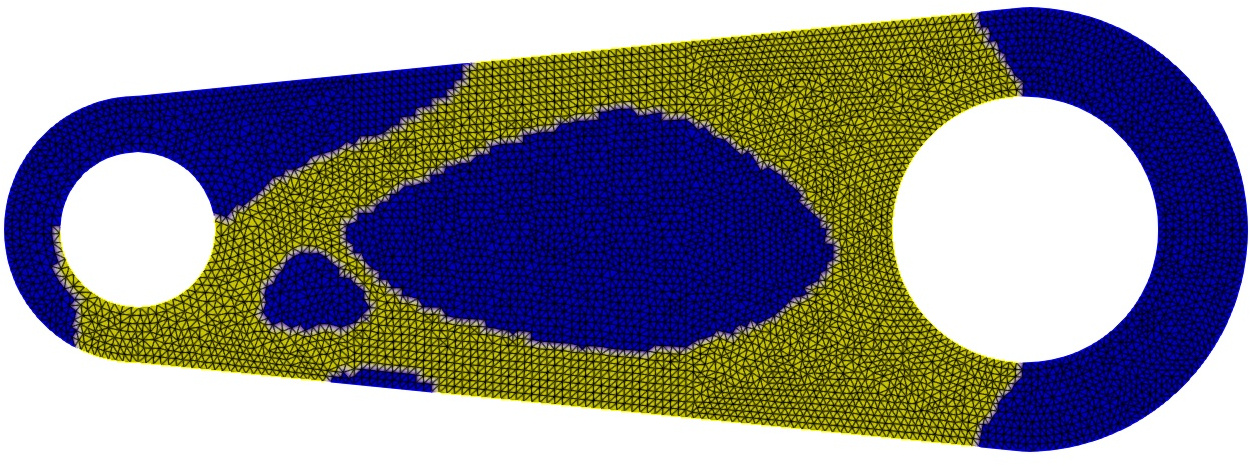}
		\caption{Test case of Section~\ref{sec:wrench} - Indicator function $\chi_{\varphi, h}^{\tt k, \rm el}$ superimposed to the level set-fitted mesh, for $\#\mathcal{T}_h^{\rm \, ls}=1000$ (top row, left) and $2000$ (top row, right); indicator function of a structure  yielded by a continuous Galerkin-based topology optimization, superimposed to the unstructured triangular mesh with 
			$N_{w, \rm ls}^c$ (mid row, left), $N_{w, \rm ls}^f$ (mid row, right), $N_{w, \rm el}^c$ (bottom row, left), $N_{w, \rm el}^f$ (bottom row, right) dofs.}
		\label{fig:cantilever_wrench}
	\end{figure}

	The last test case is borrowed from~\cite{talischi2012} and is meant to emphasize the versatility of the approach, even in the presence of complex domains. In particular, we aim at optimizing the configuration in Figure~\ref{fig:wrench_mesh}, where homogeneous Dirichlet boundary conditions are applied on the inner right circle (blue-highlighted) and a downward traction, $\bm g = [0, -1 \ \mathbb{I}_{[-0.175, 0]}(y)]^T$, is exerted on the inner left circle (red-highlighted).
	The remaining physical parameters are $E^0 = 1$ and $\nu^0 = 0.3$. \\
	We set {\tt kStart} $= 75$, {\tt kCut} $= 1$, $\Delta t = 5$e$-3$, $\alpha = 0.4$, $\tau = 1$e$-2$, while $\varphi^0$ is equal to $1$ outside of a circular hole of radius $0.25$ and centered at $(1, 0)$ (green-highlighted in Figure~\ref{fig:wrench_mesh}). We run the algorithm for two level set grids $\mathcal{T}_h^{\rm \, ls}$, consisting of $1000$ and $2000$ polygonal elements.\\ Results are shown in Figure~\ref{fig:cantilever_wrench} (first row), where the indicator function and the corresponding fitted mesh are provided. The output topology is identical regardless of the mesh resolution, and presents two holes connecting the portions of the boundary where the structure is clamped and loaded. The fitted grids consist of $1138$ and $2186$ elements, the mesh cardinality increment being below $15\%$ in both cases.
	
	The optimized structures are compared further with the result of a non-fitted level set-based topology optimization algorithm, relying on a standard continuous Galerkin setting on affine triangular finite elements~\cite{yamada2010topology}. For the sake of comparison, we select unstructured triangular meshes so that the associated discrete spaces have the same number of degrees of freedom (dofs) as the polygonal grids, while all the other parameters are kept the same. 
	In particular, since the level set-fitted approach involves the two meshes $\mathcal{T}_h^{\rm \, ls}$ and $\mathcal{T}_h^{\rm el}$, in contrast to a standard (non-fitted) topology optimization pipeline that resort to a single grid, we perform the comparison by considering both the tessellations used for the level set evolution and the linear elasticity equation approximation, 
	for $\#\mathcal{T}_h^{\rm \, ls}=1000$ and $2000$. Thus,
	recalling from Section~\ref{sec:fittedmesh_method} that space $V_h$ has $3$ dofs per element for $p = 1$,
	we deal with four meshes characterized by a number of dofs equal to
	$N_{w, \rm ls}^c = 3 \cdot 1000$, $N_{w, \rm el}^c = 3 \cdot 1138$, $N_{w, \rm ls}^f = 3 \cdot 2000$ and $N_{w, \rm el}^f = 3 \cdot 2186$ for $V_h$, associated with the coarse and the fine initial level set and level set-fitted tessellation, respectively.\\
	The cross-comparison among the optimized structures highlights that the topology delivered by the different approaches is essentially the same and independent of the selected mesh. However, it is evident that 
	the continuous Galerkin-based algorithm with unstructured meshes delivers layouts characterized by a jagged boundary (see, e.g., the mid/bottom row, left panels), while 
	the level set-fitted methodology yields structures with very regular contours, 
	despite the number of dofs remains essentially invariant. 
	Finally, the mesh cardinality plays a role in the contour description when resorting to a non-fitted approach (compare the mid/bottom, left with the mid/bottom, right wrenches). This is not the case of the level set-fitted method where the different cardinality turns out to be less impactful.
	
	\section{Conclusions and future developments}
	\label{sec:conclusions}
	In this paper, we present a methodology to locally modify a polygonal mesh so that the resulting tessellation is fitted to the zero-isoline of a level set function. The proposed approach is general, namely it can be combined with any discretization scheme. In this paper, we focus on the PolyDG approximation due to the intrinsic robustness in managing meshes including arbitrarily-shaped elements. The level set-fitted scheme thus settled is verified through a benchmark test case in linear elasticity and then integrated in a minimum compliance topology optimization workflow. 
	The main findings of the novel fitting approach are:
	\begin{itemize}
		\item the fitted meshes allow us to track crisply material discontinuities and structural interfaces, yielding more accurate results when compared with generic (non-fitted) meshes, especially when dealing with coarse grids;
		\item the proposed procedure turns out to be very cheap computationally since it does not require any local or global remeshing operation to improve the mesh quality;
		\item the delivered meshes guarantee an accurate finite element analysis (see Section~\ref{sec:lshape_validation}, for an instance), and requires only a slight mesh cardinality increment in the level set-fitting phase; 
		\item the topology optimization with level set-fitted grids delivers structures with smooth boundaries and sharp material/void alternation.
	\end{itemize}
	
	Future extensions of this work include:
	the generalization of the proposed topology optimization workflow to the control of pointwise quantities (e.g., in stress-based or compliant mechanism optimizations) or to a manufacturing-aware process (e.g., taking into account overhangs or the printing direction); 
	the use of an agglomeration strategy to suitably reduce the tessellation cardinality in the fitted mesh; the optimization and the parallelization of the code to deal with $3$-dimensional settings, where the computational benefits led by this approach are expected to be even more noteworthy.
	
	\section*{Acknowledgments}
	This research is part of the activities of MUR - grant Dipartimento di Eccellenza 2023-2027.
	All the authors gratefully acknowledge the PRIN research grant n.20204LN5N5 {\it Advanced Polyhedral Discretisations of Heterogeneous PDEs for Multiphysics Problems}, and the INdAM–GNCS membership.
	PFA and MV thank the PRIN research grant n.201744KLJ2. PFA and NP have also been partially supported by ICSC—Centro Nazionale di Ricerca in High Performance Computing, Big Data, and Quantum Computing funded by European Union—NextGenerationEU. NF acknowledges the INdAM–GNCS 2023 Project {\it Algoritmi efficienti per la gestione e adattazione di mesh poligonali}.
	
	
	
	\bibliography{manuscript}
	
	
\end{document}